\documentclass[format=acmsmall, review=false, screen=true]{acmart}

\usepackage[T1]{fontenc} % optional
\usepackage[utf8]{inputenc}
\usepackage{amsmath}
\usepackage{bm} % optional
\usepackage{graphicx}
\usepackage{tikz}
\usepackage{tcolorbox}
\usepackage{pdflscape}
\usepackage{rotating}
\usepackage{fancyhdr} 
\usepackage{color, colortbl}
\usepackage{longtable}
\usepackage{multirow}
\usepackage{enumitem}
\usepackage{fontawesome}
\usepackage{balance}
\usepackage{tablefootnote}
\usepackage{colortbl}
\tcbuselibrary{skins}
\tcbuselibrary{listingsutf8}
\usepackage[export]{adjustbox}
\usepackage{framed}
\usepackage{arydshln}
\newcolumntype{M}[1]{>{\centering\arraybackslash}m{#1}}
\newcolumntype{N}{@{}m{0pt}@{}}
\usepackage{hyperref}
\newcommand{\commentword}[1]{}

\hypersetup
{
 colorlinks = true, %Colours links instead of ugly boxes
 urlcolor = blue, %Colour for external hyperlinks
 linkcolor = blue, %Colour of internal links
 citecolor = blue %Colour of citations
}

\usepackage{graphicx}
\usepackage{svg}
\usepackage{subfig}
\usepackage{array}
\definecolor{mybeige}{HTML}{FFF7F3}
\definecolor{myoffwhite}{HTML}{F1F1F1}
\definecolor{mydarkpurple}{HTML}{49006A}
\definecolor{mypurple}{HTML}{99017B}
\definecolor{mydarkpink}{HTML}{E23E99}
\definecolor{mypink}{HTML}{F767A1}
\definecolor{mypink2}{HTML}{F769A1}
\definecolor{mylightpink}{HTML}{F994B1}
\definecolor{mysalmon}{HTML}{FCC8C3}
\definecolor{mylightsalmon}{HTML}{FBBABD}
\definecolor{lightgray}{rgb}{0.9, 0.9, 0.9}  % Light gray color for table header
\definecolor{skyblue}{rgb}{0.529, 0.808, 0.922}   % Sky blue color for highlights

\AtBeginDocument{%
  }

\setcopyright{acmlicensed}
\copyrightyear{2025}
\acmYear{2025}
\acmDOI{XXXXXXX.XXXXXXX}

\acmJournal{TOSEM}
\acmVolume{0}
\acmNumber{0}
\acmArticle{0}
\acmMonth{0}

\begin{document}

\title{Using LLMs in Generating Design Rationale for Software Architecture Decisions}

\author{Xiyu Zhou}
\email{xiyuzhou@whu.edu.cn}
\affiliation{%
  \institution{School of Computer Science, Wuhan University}
  \city{Wuhan}
  \country{China}
}

\author{Ruiyin Li}
\authornote{Corresponding author}
\email{ryli_cs@whu.edu.cn}
\affiliation{%
  \institution{School of Computer Science, Wuhan University}
  \city{Wuhan}
  \country{China}
}

\author{Peng Liang}
\authornote{Corresponding author}
\email{liangp@whu.edu.cn}
\affiliation{%
  \institution{School of Computer Science, Wuhan University}
  \city{Wuhan}
  \country{China}
}

\author{Beiqi Zhang}
\email{zhangbeiqi@whu.edu.cn}
\affiliation{%
  \institution{School of Computer Science, Wuhan University}
  \city{Wuhan}
  \country{China}
}

\author{Mojtaba Shahin}
\affiliation{%
  \institution{RMIT University}
  \country{Australia}
}
\email{mojtaba.shahin@rmit.edu.au}

 \author{Zengyang Li}
\affiliation{%
  \institution{School of Computer Science, Central China Normal University}
  \country{China}
}
\email{zengyangli@ccnu.edu.cn}

\author{Chen Yang}
\email{yangchen@szpu.edu.cn}
\affiliation{%
  \institution{School of Artificial Intelligence, Shenzhen Polytechnic University}
  \city{Shenzhen}
  \country{China}
}
% \author{Peng Liang}
% \email{liangp@whu.edu.cn}
% \authornote{Corresponding author}
% \affiliation{%
%   \institution{School of Computer Science, Wuhan University}
%   \city{Wuhan}
%   \country{China}
% }

%% The "author" command and its associated commands are used to define the authors and their affiliations. Of note is the shared affiliation of the first two authors, and the "authornote" and "authornotemark" commands used to denote shared contribution to the research.
% \affiliation{%

\renewcommand{\shortauthors}{Zhou et al.}

\begin{abstract}
Design Rationale (DR) for software architecture decisions refers to the reasoning underlying architectural choices, which provides valuable insights into the different phases of the architecting process throughout software development. However, in practice, DR is often inadequately documented due to a lack of motivation and effort from developers. With the recent advancements in Large Language Models (LLMs), their capabilities in text comprehension, reasoning, and generation may enable the generation and recovery of DR for architecture decisions. In this study, we evaluated the performance of LLMs in generating DR for architecture decisions. First, we collected 50 Stack Overflow (SO) posts, 25 GitHub issues, and 25 GitHub discussions related to architecture decisions to construct a dataset of 100 architecture-related problems. Then, we selected five LLMs to generate DR for the architecture decisions with three prompting strategies, including zero-shot, chain of thought (CoT), and LLM-based agents. With the DR provided by human experts as ground truth, the Precision of LLM-generated DR with the three prompting strategies ranges from 0.267 to 0.278, Recall from 0.627 to 0.715, and F1-score from 0.351 to 0.389. Besides, 64.45\% to 69.42\% of the arguments of DR not mentioned by human experts are also helpful, 4.12\% to 4.87\% of the arguments have uncertain correctness, and 1.59\% to 3.24\% of the arguments are potentially misleading. \textcolor{black}{To further understand the trustworthiness and applicability of LLM-generated DR in practice, we conducted semi-structured interviews with six practitioners. Based on the experimental and interview results, we discussed the pros and cons of the three prompting strategies, the strengths and limitations of LLM-generated DR, and the implications for the practical use of LLM-generated DR.}
\end{abstract}

\begin{CCSXML}
<ccs2012>
 <concept>
  <concept_id>00000000.0000000.0000000</concept_id>
  <concept_desc>Do Not Use This Code, Generate the Correct Terms for Your Paper</concept_desc>
  <concept_significance>500</concept_significance>
 </concept>
 <concept>
  <concept_id>00000000.00000000.00000000</concept_id>
  <concept_desc>Do Not Use This Code, Generate the Correct Terms for Your Paper</concept_desc>
  <concept_significance>300</concept_significance>
 </concept>
 <concept>
  <concept_id>00000000.00000000.00000000</concept_id>
  <concept_desc>Do Not Use This Code, Generate the Correct Terms for Your Paper</concept_desc>
  <concept_significance>100</concept_significance>
 </concept>
 <concept>
  <concept_id>00000000.00000000.00000000</concept_id>
  <concept_desc>Do Not Use This Code, Generate the Correct Terms for Your Paper</concept_desc>
  <concept_significance>100</concept_significance>
 </concept>
</ccs2012>
\end{CCSXML}

\ccsdesc[500]{Software and its engineering~Designing software}
\ccsdesc[500]{General and reference~Empirical studies}

\keywords{Design Rationale, Architecture Decision, Large Language Model, Prompt Engineering, LLM-based Agent}



\maketitle

\section{Introduction}
The software architecture of a system consists of the structures required to understand the system, including its software elements, their relationships and properties \cite{bass2021software}. When making architecture decisions, software architects need to consider multiple factors like the application domain, architectural styles and patterns, Commercial Off-The-Shelf (COTS) components, and other infrastructure choices required to meet system requirements \cite{jansen2005software}. Design Rationale (DR) explains the reasoning behind architecture decisions, encapsulating the architecture knowledge and thought processes that support the resulting design \cite{tang2006asurvey}. A comprehensive DR can support various development activities, such as change impact analysis or a major redesign. With documented DR, architecture decisions can be more easily revisited and assessed by software architects \cite{davide2013value}.

% However, several previous studies highlight a significant gap in capturing and understanding DR in real-world software development \cite{tyree2005architecture, bosch2004software}. While the importance of documenting the DR behind architecture decisions is widely acknowledged, practitioners frequently underestimate its value and face challenges in effectively capturing and recording it. Consequently, many software engineers admit to struggling to understand the reasoning behind architecture decisions. \cite{tang2006asurvey}.

Although the importance of DR in architecture design has been widely acknowledged, it is often inadequately recorded and detailed during software development \cite{tyree2005architecture, bosch2004software}. Tang \textit{et al.}~\cite{tang2006asurvey} conducted a survey to investigate the oversight of DR by developers (e.g., ``not aware of'' and ``no time/budget'') and the challenges of capturing DR (e.g., ``no suitable tool''). Besides, extra efforts are required when architects need to capture DR \cite{capilla2008effort}. The absence of DR hinders developers from understanding the reasons behind architecture decisions, thus violating the architecture decisions made in the design phase during the implementation and increased maintenance costs \cite{tang2007rationale}, which reveals a critical gap in the DR capture process during software architecting.

The advancement of Large Language Models (LLMs) offers promising potential to address this gap by assisting in generating DR. LLMs, powered by deep learning technologies, have revolutionized Natural Language Processing (NLP) by demonstrating advanced language understanding capabilities, including syntax, semantics, and pragmatics \cite{brown2020language}. LLMs' ability to generate diverse content stems from extensive parameters and training on large-scale datasets, enabling sustained performance improvements across a wide range of downstream tasks \cite{wei2022emergent}. Previous studies (e.g., \cite{kabir2024stack, jin2024mare, widjojo2023addressing}) have demonstrated LLMs' utility in addressing Software Engineering (SE) challenges, highlighting their potential to support developers in various SE tasks. \textcolor{black}{In particular, by leveraging the natural language comprehension and reasoning capabilities of LLMs \cite{jason2022chain}, software architects can efficiently use LLMs to assist in generating and recovering DR of architecture decisions.} Meanwhile, the accuracy and quality of DR generated by LLMs are yet to be explored. Besides, according to the study by Soliman \textit{et al.} \cite{soliman2025do}, who employed a zero-shot method to query GPT-3.5 regarding the DR of the Hadoop Distributed File System (HDFS)\footnote{\url{https://hadoop.apache.org/docs/stable/hadoop-project-dist/hadoop-hdfs/HdfsDesign.html}}, GPT-3.5 exhibited moderate Recall but relatively low Precision against a predefined ground truth for DR derived from architecture issues and related documents on HDFS. Therefore, how to effectively guide LLMs to ensure the generation of high-quality DR remains a critical challenge that needs further investigation.
% However, little is known about the accuracy and quality of DR generated by LLMs. In addition, since architecture decisions often involve multiple layers of considerations (e.g., design patterns and technology choices \cite{bass2012software}), inferring DR can be quite complex. Therefore, how to effectively leverage existing technologies (e.g., prompt engineering \cite{liu2023pre} and LLM-based agents \cite{liu2024agent}) to guide LLMs in generating high-quality and accurate DR is also a critical challenge that needs to be addressed.

Our study \textbf{aims} to explore and validate the feasibility of leveraging LLMs to uncover the DR that underpins the software architecture decisions. By achieving this goal, we seek to enhance the understanding of legacy software systems for developers and stakeholders, thereby ultimately facilitating more effective maintenance and evolution of software systems.

In this paper, we conducted an empirical study to evaluate the DR of architecture decisions generated by LLMs. We first constructed a dataset containing architecture-related problems from software development scenarios. Specifically, the dataset is composed of 100 data points related to architecture decisions, including 50 posts from SO, along with 25 discussions and 25 issues from GitHub. We then chose five widely used LLMs for the evaluation, including both popular open-source and proprietary models: gpt-3.5-turbo, gpt-4-0613, gemini-1.0-pro, llama3-8B, and mistral-7B. To generate DR of architecture decisions, we employed three distinct prompting strategies: zero-shot, chain of thought (CoT), and LLM-based agents. For zero-shot and CoT methods, we designed tailored prompt templates, while for LLM-based agents, we developed a multi-agent system consisting of five LLM-based agents that collaborate to generate DR. Finally, we evaluated the DR generated by the five LLMs and the three prompting strategies using both accuracy metrics for quantitative analysis and IHUM-category for qualitative analysis. \textcolor{black}{Specifically, for quantitative analysis, we employed Precision, Recall, and F1-score to evaluate the generated DR, using the rationale provided by human experts as ground truth. For qualitative analysis, we introduced the IHUM-category to classify the argument points in the LLM-generated DR into four categories, allowing us to further assess the arguments not mentioned by the human experts. The detailed definition of the IHUM-category is provided in Table~\ref{tab:IHUMcategory}.} \textcolor{black}{Based on the results of our experiments, we conducted semi-structured interviews with six practitioners to understand the trustworthiness and applicability of LLM-generated DR in practice. From the perspective of practitioners, we explored their trust level in LLM-generated DR, the information that needs to be provided to LLMs to enhance their trust, their willingness to use LLM-generated DR in practice, suitable scenarios for its application, as well as the challenges and limitations associated with applying LLM-generated DR.}

The main \textbf{contributions} of this work are that:
\begin{itemize}
    \item We constructed and publicly released a dataset containing 100 architecture-related problems from software development scenarios on SO and GitHub \cite{dataset}.
    \item We developed tailored prompt templates for zero-shot and CoT methods, and designed a multi-agent system consisting of five LLM-based agents to generate DR for architecture decisions.
    \item We evaluated the accuracy and quality of the DR generated by LLMs through both accuracy metrics and IHUM-category. 
    \item \textcolor{black}{We conducted semi-structured interviews with practitioners to understand the trustworthiness and applicability of LLM-generated DR in practice.}
    \item \textcolor{black}{We discussed the advantages and disadvantages of the three prompting strategies, the strengths and limitations of LLM-generated DR in comparison to that provided by human experts, as well as the implications for practical use of LLM-generated DR.}
\end{itemize}
% \begin{itemize}
%     \item This is the first study to systematically investigate fairness concerns in AI-based apps on a large scale from the users' perspective. To facilitate replication and further study, we make our scripts and collected data available online \cite{}.
%     \item We establish a ground-truth dataset consisting of 1,132 fairness reviews and 1,473 non-fairness reviews. Based on the dataset, we train ML and DL models that can accurately distinguish fairness from non-fairness reviews.
%     \item We propose a clustering and summarization approach to discover six fine-grained fairness concerns of users. Besides, we summarize six root causes of fairness concerns from the app owners' perspective.
%     % \item We make the source code and experimental data available in our replication package~\cite{}.
% \end{itemize}

\textbf{Paper Organization}: Section~\ref{sec:background} introduces the background knowledge relevant to this study. Section~\ref{sec:methodology} presents the Research Questions (RQs) and the research process. Section~\ref{sec:results} presents the evaluation results of LLM-generated DR with their interpretation. Section~\ref{sec:discussion} discusses the implications based on the research results. Section~\ref{sec:validity} clarifies the potential threats to the validity of this study. Section~\ref{sec:relatedwork} reviews the related work. Finally, Section~\ref{sec:conclusion} concludes this work and outlines future research directions.

\section{Background}\label{sec:background}
\subsection{Definition of Design Rationale}
Typically, design rationale in software development refers to the underlying reasons for making design decisions \cite{moran2020design}. However, no standardized definition of DR in software design has been established, resulting in subtle variations in its interpretation across different contexts \cite{moran2020design}. For example, Rogers \textit{et al.} \cite{rogers2015using} define that DR consists of the decision problem, alternative solutions, and arguments for or against the solutions. Besides, Zhao \textit{et al.} \cite{zhao2024drminer} defined DR as a combination of a solution and its corresponding supporting or opposing arguments. 

We provided the precise interpretation of DR in this study. Our dataset is derived from the posts on SO, as well as the issues and discussions on GitHub. Each data point includes a specific architecture-related problem  \(\bm{P}\); an architecture decision \(\bm{D}\); and its corresponding design rationale \(\bm{DR}\). Design rationale (\(\bm{DR}\)) is composed of a set of arguments \(\{\bm{A}_1, \bm{A}_2, \ldots, \bm{A}_n\}\), where each argument corresponds to one particular perspective for architecture design (\(\bm{D}\)), such as advantages, disadvantages, and trade-offs. The relationship between architecture problem \(\bm{P}\), architecture decision \(\bm{D}\), and design rationale \(\bm{DR}\) in the dataset is presented in Equation~\eqref{eq:DRdefinition}. \textcolor{black}{Equation~\eqref{eq:DRdefinition} specifies which data items should be extracted from SO posts, GitHub issues, and GitHub discussions to construct our dataset. Accordingly, the data items D1--D3 presented in Table~\ref{tab:dataitem} correspond to \(\bm{P}\), \(\bm{D}\), and \(\bm{DR}\), respectively. Moreover, Equation~\eqref{eq:DRdefinition} defines DR as comprising multiple argument points. By decomposing DR into individual arguments, we can evaluate these arguments using both accuracy metrics and the IHUM-category.}

\begin{equation}
\label{eq:DRdefinition}
\{(P, D, DR) \mid DR = \{A_{1}, A_{2}, \dots, A_{n}\}, n \geq 1 \}
\end{equation}

\subsection{Prompt Engineering of LLMs}
Prompt engineering has emerged as an essential technique for enhancing the capabilities of pre-trained LLMs \cite{sahoo2024systematic}. By designing prompts to guide the output, LLMs can adapt to various tasks and domains without requiring modifications to their parameters~\cite{widyasari2024codeagent, deljouyi2024leveraging, delile2023evaluating}. The field of prompt engineering encompasses a wide range of techniques tailored to optimize LLM performance. 

In this study, we focused on two prompting methods: zero-shot and CoT. Zero-shot is a foundational prompting technique where the task is directly presented to LLMs without requiring additional examples. This approach enables LLMs to perform tasks by leveraging its pre-training knowledge \cite{brown2020language}. We employed the zero-shot method to evaluate the performance of LLMs in generating DR using the basic prompt. On the other hand, CoT is a technique that guides LLMs through a structured sequence of reasoning steps. Specifically, the template of the CoT method instructs LLMs to generate more thoughtful and deeply reasoned answers by following a series of steps. Wei \textit{et al.} \cite{jason2022chain} demonstrated that CoT \commentword{significantly} improves performance in math and commonsense reasoning tasks. Given that DR often involves complex reasoning and trade-offs, we employed CoT in order to help LLMs produce more carefully considered outputs. The templates used for zero-shot and CoT methods are presented in Section~\ref{sec:methodology}.

\subsection{LLM-based Agents}
An intelligent agent is an automated entity capable of perceiving, making decisions, and taking actions \cite{wooldridge1995the}. With the advent of LLMs, researchers and practitioners have increasingly focused on developing LLM-based autonomous agents, leading to the emergence of LLM-based agent frameworks (e.g., BabyAGI \cite{babyagi} and MetaGPT \cite{hong2023metagpt}). These agent frameworks typically comprise multiple autonomous intelligent agents, each capable of independently performing tasks by interacting, communicating, and collaborating with other agents. Compared to prompt engineering, multi-agent systems offer more sophisticated agent interactions and coordination mechanisms. By leveraging a variety of tools, LLM-based agent systems can further enhance their capabilities, enabling the systems to collectively address more complex and dynamic challenges. Several LLM-based agent systems have been developed to tackle different tasks in SE, such as code generation \cite{bairi2024codeplan}, code review \cite{hu2023large}, and requirements engineering \cite{jin2024mare}. However, there is currently no LLM-based agent systems specifically designed for generating DR for architecture decisions. Considering the complexity of DR, we designed a multi-agent system consisting of five LLM-based agents, building on the MetaGPT framework \cite{hong2023metagpt}. The details of this LLM-based agent system are presented in Section~\ref{sec:methodology}.

\section{Methodology}\label{sec:methodology}
\begin{figure}[h]
    \centering
    \makebox[\textwidth]{%
        \includegraphics[width=1.0\textwidth]{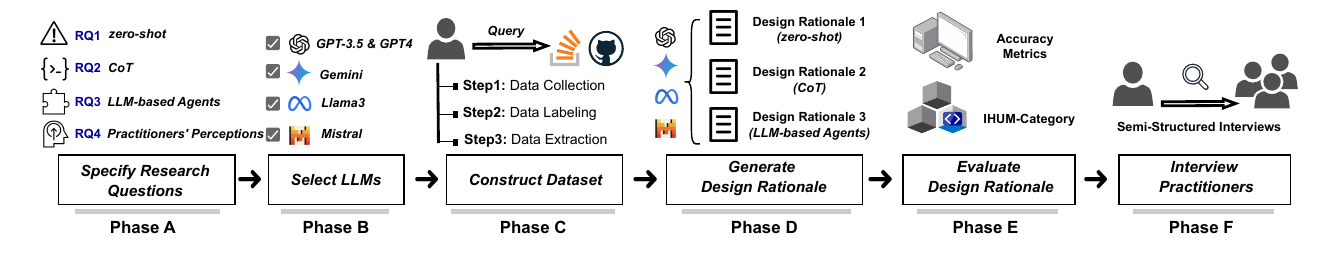}
    }
    \caption{\textcolor{black}{Overview of the research process}}
    \label{fig:researchprocess}
\end{figure}

In this section, we introduce the methodology employed in our research in terms of research questions (Section~\ref{subsec:RQs}), selection of LLMs (Section~\ref{subsec:LLMselection}), dataset construction (Section~\ref{subsec:datasetconstruction}), design rationale generation (Section~\ref{subsec:rationalegeneration}), evaluation methods (Section~\ref{subsec:evaluation}), \textcolor{black}{and interviews with practitioners} (Section~\ref{subsec:interviews_steps}). The overview of the research process is presented in Fig.~\ref{fig:researchprocess}.

\subsection{Research Questions}
\label{subsec:RQs}
Our goal is to uncover and evaluate the capability of LLMs in generating DR for architecture decisions, \textcolor{black}{as well as practitioners’ perceptions of LLM-generated DR in practice.} To this end, we formulated the following \textcolor{black}{four} Research Questions (RQs):

\begin{tcolorbox}[colback=gray!20, colframe=gray]
\textbf{RQ1: How effective are LLMs in generating and recovering design rationale for architecture decisions?}
\end{tcolorbox}

\noindent \textbf{\textit{Rationale}}: RQ1 aims to explore the feasibility and capability of using LLMs in generating DR for architecture decisions. DR justifies architecture choices and helps long-term system maintainability by providing context for software engineers to understand architecture decisions \cite{tang2007rationale}. LLMs can potentially automate the DR generation, helping recover and preserve the reasons behind architecture decisions.

\begin{tcolorbox}[colback=gray!20, colframe=gray]
\textbf{RQ2: Does the CoT technique affect or enhance the capability of LLMs to generate and recover design rationale for architecture decisions?}
\end{tcolorbox}

\noindent \textbf{\textit{Rationale}}: 
CoT, involving a series of intermediate reasoning steps, \commentword{significantly} enhances the ability of LLMs to perform complex reasoning \cite{jason2022chain}. Since the CoT method enables step-by-step reasoning, it can potentially instruct LLMs to engage in more comprehensive thinking and make more refined trade-offs for architecture decisions. Therefore, we plan to explore the ability of the CoT method to enhance the quality of LLM-generated DR for architecture decisions.

\begin{tcolorbox}[colback=gray!20, colframe=gray]
\textbf{RQ3: Do LLM-based agents affect or enhance the capability of LLMs to generate and recover design rationale for architecture decisions?}
\end{tcolorbox}

\noindent \textbf{\textit{Rationale}}: 
LLM-based agents have been successfully applied in the field of SE, demonstrating significant effectiveness in supporting and facilitating various SE tasks \cite{liu2024large}. By collaborating in specific roles, LLM-based agents are capable of performing complicated tasks such as reasoning, gathering information, and reviewing output. However, it remains unclear whether LLM-based agents can effectively generate DR for architecture decisions. Therefore, we designed a multi-agent system to explore whether collaboration among LLM-based agents can further enhance the ability of LLMs to generate DR for architecture decisions.

\begin{tcolorbox}[colback=gray!20, colframe=gray]
\textbf{\textcolor{black}{{RQ4: How do practitioners perceive LLM-generated DR of architecture decisions in practice?
}}}
\end{tcolorbox}

\noindent \textcolor{black}{\textbf{\textit{Rationale}}: 
If the performance of LLMs in generating DR for architecture decisions is acceptable, their use for automatic DR generation could provide practical value for software architects. However, the perception of LLM-generated DR from the practitioners’ perspective remains unclear. To address RQ4, we conducted semi-structured interviews to investigate how practitioners evaluate the DR generated by LLMs in terms of trustworthiness and applicability.}

\subsection{Selection of LLMs}
\label{subsec:LLMselection}
As presented in Table~\ref{tab:selectedLLM}, we selected five mainstream open-source and closed-source LLMs in this study. \textcolor{black}{These LLMs exhibit high performance in natural language understanding and reasoning, and have been widely applied in many SE tasks (e.g., \cite{kabir2023answers, Sungmin2024evaluating, de2023lessons, zhang2024using}). Besides, the knowledge cut-off date of LLMs is also an important factor in our model selection. Since SO posts, GitHub issues and discussions, are publicly available, we collected data generated after the knowledge cut-off dates of all employed LLMs to prevent data leakage issues \cite{balloccu2024leak}. Note that selecting an LLM with a more recent cut-off date may result in a smaller dataset, potentially impacting the analysis of experimental results. For instance, if we choose GPT-4o, one of the most advanced models from OpenAI~\cite{gpt-4o}, only 28 architecture-related posts from SO, which were created after the knowledge cut-off date of GPT-4o, would remain available.} \textcolor{black}{Therefore, after considering the trade-off between model recency and dataset size, we ultimately selected the five LLMs listed in Table~\ref{tab:selectedLLM}.}
\textcolor{black}{Among the five selected models, llama3-8B is the only LLM released in 2024, while the others were all released in 2023. The time gap between llama3-8B and the earliest model, gpt-3.5-Turbo, spans one year and one month. We selected llama3-8B due to its substantial improvements over previous versions in the LLaMA series. More importantly, its knowledge cut-off date (March 2023) is earlier than that of gemini-1.0-pro (April 2023), ensuring that it does not constrain the dataset size.}
% As a result, we did not utilize GPT-4o in this study. However, we will plan to employ this model in our future research.

\begin{table}[ht]
    \centering
    \small
    \caption{\textcolor{black}{Five selected LLMs to generate DR of architecture decisions}}
    \begin{tabular}{p{1cm}p{6.1cm}>{\centering\arraybackslash}p{1.7cm}>{\centering\arraybackslash}p{1.75cm}>{\centering\arraybackslash}p{1.85cm}}
        \toprule
        \rowcolor{lightgray} \textbf{Model} & \textbf{Description} & \textbf{Cut-off date} & 
        \textbf{Release date} &
        \textbf{Open-source} \\ \hline
        gpt-3.5-turbo & gpt-3.5-turbo is an LLM from OpenAI's GPT series. It can understand and generate natural language and code, excelling in tasks like conversation generation, text completion, and question answering. While optimized for chat applications, it also effectively handles non-chat tasks. & 2021-09 & \textcolor{black}{2023-03} & No \\ \hline
        gpt-4-0613 & gpt-4-0613 is an advanced multimodal LLM that accepts text and image inputs and produces text outputs. It offers improved accuracy and problem-solving capabilities over previous models, leveraging its extensive knowledge and reasoning skills. gpt-4-0613 is versatile, suitable for chat, text generation, and traditional completion tasks. & 2021-09 & \textcolor{black}{2023-06} & No \\ \hline
        gemini-1.0-pro & gemini-1.0-pro is an LLM developed by Google, capable of understanding and generating human language with remarkable proficiency. It can process not only text, but also understand and generate various forms of information such as images and videos. gemini-1.0-pro excels in tasks like multilingual translation, code generation, and question answering. & 2023-04 & \textcolor{black}{2023-12} & No \\ \hline
        llama3-8B & llama3-8B is an open-source LLM developed by Meta for efficient natural language processing. It excels in tasks like text understanding, generation, and conversation. With a flexible design, it operates across various platforms and is freely available for researchers and developers to modify. &2023-03 &\textcolor{black}{2024-04} & Yes \\ \hline
        mistral-7B & mistral-7B is an LLM developed by Mistral AI, designed for high-performance natural language processing tasks. It excels in text generation, summarization, and understanding, offering state-of-the-art capabilities for developers and researchers. Additionally, mistral-7B is open-source, allowing for customization and integration into various projects. &2021-08 & \textcolor{black}{2023-09} & Yes \\ \bottomrule
    \end{tabular}
    \label{tab:selectedLLM}
\end{table}

\subsection{Dataset Construction}
\label{subsec:datasetconstruction}
\begin{figure}[h]
    \centering
    \makebox[\textwidth]{%
        \includegraphics[width=0.7\textwidth]{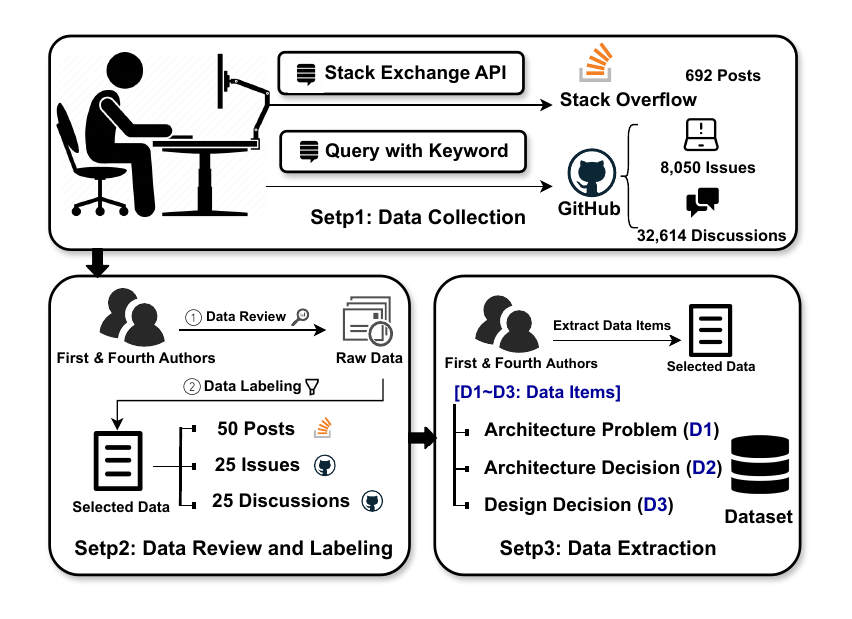}
    }
    \caption{Dataset construction regarding architecture decisions}
    \label{fig:datasetconstruction}
\end{figure}

As shown in Fig.~\ref{fig:datasetconstruction}, we constructed a dataset from SO posts, GitHub issues and discussions to facilitate and evaluate the DR generation for practical architecture decisions. To avoid potential data leakage issues, we constructed a new dataset, as the existing datasets may contain data that was used to train the LLMs used in this study as we discussed in Section~\ref{subsec:LLMselection}. Note that, in our dataset, all posts, issues and discussions were created after the cut-off dates of the five LLMs used in this study. The dataset used in this study has been provided at \cite{dataset}.

\subsubsection{Data Collection}
% For data collection, we utilized an SQL query through the Stack Exchange Data Explorer interface \cite{stackexchange}, a web-based tool that allows the execution of SQL queries on data from various Q\&A sites, including SO. We selected the general terms “architect*” (including ``architecture'', ``architecting'', and ``architectural'') to search ARPs that at least contain one answer in SO, as the research by Musengamana \textit{et al.} \cite{musengamana2023characterizing} indicates that these terms can help to capture more relevant posts compared to other general keywords (e.g., ``design*''). 
For data collection, we used a keyword-based method to gather data from SO and GitHub. Specifically, we selected the general terms ``architect*'' (including ``architecture'', ``architecting'', and ``architectural'') to search posts, issues and discussions. Such selection was also used in a recent study to retrieve architecture related SO posts \cite{musengamana2023characterizing}, and it indicates that these terms are effective in capturing more relevant data compared to other general keywords like ``design*''. For SO posts, we utilized an SQL query via Stack Exchange Data Explorer interface \cite{stackexchange} to collect posts with at least one answer from SO. For GitHub issues and discussions, we collected issues marked as ``closed'' with the reason specified as ``completed'', and discussions marked as ``answered''. Note that, we did not limit our search solely to the tags of SO posts, GitHub issues and discussions, as using tags alone may omit some important data related to software architecture topics \cite{barua2014developers}. Therefore, we considered both the body and title of posts, issues and discussions during our data search.

Additionally, we limited the retrieval of the posts, issues and discussions to dates after the knowledge cut-off dates of all LLMs. As presented in Table~\ref{tab:selectedLLM}, gemini-1.0-pro has the latest cut-off date among the five LLMs we selected, that is, April 2023. Hence, we restricted the data retrieved from SO and GitHub to content after May 2023. We conducted data collection on June 18, 2024, gathering data from SO and GitHub that contained ``architect*'' in the content body, tags, or title, starting from May 1, 2023. In total, 692 SO posts, 32,614 issues, and 2,546 discussions were retrieved. 
%We provided the complete SQL query used to collect ARPs in our dataset \cite{dataset}.

\subsubsection{Data Review and Labeling}
The purpose of the data review and labeling process is to remove data entries that are not applicable to this study. The four criteria for data labeling are as follows:
\begin{itemize}
    \item \textbf{For SO posts, they must include an answer that was either accepted by the question creator or received upvotes from the SO community.} This criterion ensures that the architecture decision extracted from the post is sufficiently reasonable to address the current architecture problem.
    \item \textbf{For GitHub issues and discussions, they must come from repositories with more than 200 stars and must not stem from private projects (e.g., coding exercises).} This criterion ensures that the architecture decisions come from publicly available projects with a certain level of popularity, thereby reducing the likelihood that the extracted architecture decisions originate from toy projects.
    \item \textbf{The posts, issues, and discussions must be relevant to architecture decisions.} Although we collected data that contain ``architect*'' in tags, titles, or bodies, not all of these data entries are related to software architecture. For example, in certain SO posts, ``\textit{architecture}'' refers to ``\textit{cpu-architecture}'', which is focused on hardware rather than software architecture design.
    \item \textcolor{black}{\textbf{The posts, issues, and discussions must explicitly include an ``architecture-related problem'' \textcolor{black}{(\(\bm{P}\))}, and the corresponding ``architecture decision'' \textcolor{black}{(\(\bm{D}\))} and ``design rationale'' \textcolor{black}{(\(\bm{DR}\))}}. \textcolor{black}{The relationship among \(\bm{P}\), \(\bm{D}\), and \(\bm{DR}\), as defined in Equation~\ref{eq:DRdefinition}, serves as the basis for our data extraction, ensuring that the required information (i.e., D1--D3 presented in Table~\ref{tab:dataitem}) could be retrieved from each data entry.}}
\end{itemize}

The formal data labeling was conducted by the first author. Prior to this, the first and fourth authors randomly selected 25 SO posts, 25 GitHub issues, and 25 GitHub discussions for a pilot labeling process to minimize personal bias and ensure adherence to the data labeling criteria. The inter-rater reliability between the two authors was measured by the Cohen's Kappa coefficient \citep{jacob1960coefficient}, yielding values of 0.834. This result indicates an almost perfect agreement between the two authors \cite{landis1977theMO}. The second and third authors were then involved to resolve discrepancies between the pilot results of the first and fourth authors and reach a consensus on the four criteria for data labeling. Subsequently, the first author conducted the formal data labeling according to the four criteria. \textcolor{black}{For SO posts, the first author reviewed and labeled all 692 posts, and identified 50 architecture-related posts for our research.} \textcolor{black}{During the data labeling process, 83 posts were excluded based on Criterion~1, whereas 358 and 204 posts were excluded based on Criteria~3 and 4, respectively.} \textcolor{black}{Given the substantial number of collected GitHub issues and discussions, and to maintain a balanced data volume between SO and GitHub, the two data sources, we decided to select 25 issues and 25 discussions that satisfied the inclusion criteria. Consequently, the final dataset comprises 100 entries, evenly split between the two data sources: 50 from SO and 50 from GitHub.} \textcolor{black}{The 25 issues and 25 discussions were obtained after the first author manually reviewed and labeled 586 GitHub issues and 784 GitHub discussions, respectively. A total of 114 issues and 21 discussions were excluded based on Criterion~2, 272 issues and 498 discussions were excluded based on Criterion~3, and 175 issues and 240 discussions were excluded based on Criterion~4. To assess whether the reviewed GitHub issues and discussions are representative of the collected data, we conducted statistical estimation based on the review size and the total number of collected data entries \cite{israel1992determining}. As a result, the margin of error was calculated to be 4.01\% for the GitHub issues and 2.91\% for the GitHub discussions at the 95\% confidence level.} The data labeling results were reviewed multiple times by the first, second, third, and fourth authors. Discrepancies were addressed through discussions to reach a consensus. Specifically, the first author presented the reasons for the inclusion or exclusion of certain data. The four authors then analyzed and discussed the reasons based on the criteria for data labeling, ultimately reaching an agreement. The pilot and formal data labeling results are recorded in our dataset \cite{dataset}.

\textcolor{black}{Table~\ref{tab:architecture_decision_taxonomy} presents the taxonomy of eight types of architecture decisions with examples from our dataset, developed through the constant comparison method~\cite{stol2016grounded}. Within the 100 architecture decisions collected from our dataset, \textit{Implementation Decisions} (25\%) are observed with the highest frequency, followed by \textit{Technology Decisions} (15\%) and \textit{Integration Decisions} (14\%). \textit{Architecture Pattern Decisions} (12\%), \textit{Data Decisions} (12\%), and \textit{Component Decisions} (11\%) occur less frequently, whereas \textit{Design Pattern Decisions} (6\%) and \textit{Deployment Decisions} (5\%) are the least frequent.}

\begin{table}[htbp]
  \centering
  \caption{\textcolor{black}{Taxonomy of architecture decisions}}
  \label{tab:architecture_decision_taxonomy}
  \color{black}{
  \begin{tabular}{l p{8cm} c}
    \toprule
    \textcolor{black}{\textbf{Type}} & \textcolor{black}{\textbf{Example}} & \textcolor{black}{\textbf{Count}}  \\
    \midrule
    Implementation Decision & \textit{If you want to auto-register the dependencies, use a different library or implement it yourself - e.g. take the domain Assembly, enumerate all types that end with ``Query'' and register them as scoped} (\href{https://web.archive.org/web/20250622090505/https://stackoverflow.com/questions/77843622/dependency-injection-on-an-n-tier-solution-remove-boilerplate}{SO~\#77843622}) &25 \\ \hline
    Technology Decision & \textit{From what I get, SignalDB should work fine with them (MPA and the Islands architecture).} (\href{https://web.archive.org/web/20250623041101/https://github.com/maxnowack/signaldb/discussions/626}{Discussion~\#626})&15\\ \hline
    Integration Decision & \textit{I suggest writing all API controllers and authentication logic in the server project. Create a models project and reference this in both the client and server projects} (\href{https://web.archive.org/web/20250622085856/https://stackoverflow.com/questions/77815059/regarding-blazor-web-app-net-8-and-migration-from-net-6}{SO \#77815059}) &14 \\ \hline
    Architecture Pattern Decision & \textit{You would have to go for another architectural style, e.g. Onion Architecture.} (\href{https://web.archive.org/web/20250623050517/https://github.com/xmolecules/jmolecules/discussions/112}{Discussion~\#112}) &12 \\ \hline
    Data Decision & \textit{Instead, we should support ref in foreign key constraint expression - both at the model and column level.} (\href{https://web.archive.org/web/20250623052141/https://github.com/dbt-labs/dbt-core/issues/8062}{Issue~\#8062}) &12 \\ \hline
    Component Decision & \textit{The second option is to pull Path.State (and all child feature States) into a separate module.} (\href{https://web.archive.org/web/20250623050023/https://github.com/pointfreeco/swift-composable-architecture/discussions/2792}{Discussion~\#2792}) &11 \\ \hline
    Design Pattern Decision &\textit{In Dart, we can use "implements" on a concrete class, so it is now my general pattern to make a concrete class for the repository, and merely ``implement'' that for my mock repository for testing.} (\href{https://web.archive.org/web/20250622084557/https://stackoverflow.com/questions/77579165/this-is-about-the-repository-layer-when-implementing-a-clean-architecture-using}{SO \#77579165})&6 \\ \hline
    Deployment Decision & \textit{Approach: maintain a map or comparable structure of prisma client instances in the nextjs app / node server instance in memory, one for each db / schema instance requested} (\href{https://web.archive.org/web/20250622092844/https://github.com/prisma/prisma/discussions/20920}{Discussion~\#2090})&5\\
    \bottomrule
  \end{tabular}
  }
\end{table}

% The data labeling process was conducted by the first author. To minimize personal bias in the formal labeling process, the first and third authors randomly \textcolor{black}{selected 25 collected SO posts, 25 GitHub issues and 25 discussions} to conduct a pilot data labeling. The inter-rater reliability between the two authors was measured by Cohen's Kappa coefficient \citep{jacob1960coefficient}, yielding values of 0.834. This result indicates an almost perfect agreement between the two authors \cite{landis1977theMO}. \textcolor{black}{Likewise, the two authors conducted the formal data labeling process according to the four criteria. For SO posts, the two authors reviewed and labeled all 692 posts, and identified 50 architecture-related posts for our research. Given the large number of issues and discussions collected from GitHub, and to maintain a balanced data volume between SO and GitHub sources, we selected the top 25 issues and top 25 discussions that meet the criteria, sorted by ``best match''. Consequently, our dataset comprises 100 entries from SO and GitHub.}
% Noted that, during both the pilot and formal data labeling, the two authors collaborated with the second author to achieve a consensus in instances of disagreement in the results. Specifically, the first author and third author presented their reasons for the inclusion or exclusion of certain data. The three authors analyzed and discussed these reasons based on the criteria for data labeling, ultimately reaching an agreement. The data labeling results are recorded in MS Excel \cite{dataset}.

\subsubsection{Data Extraction}
\textcolor{black}{According to Equation~\ref{eq:DRdefinition}, we established a set of data items for data extraction, as presented in Table~\ref{tab:dataitem}.} Data items D1 and D2 are provided as part of the prompt for LLMs and LLM-based agents to generate DR. D3 serves as a reference for evaluation in results analysis, as further elaborated in Section~\ref{subsec:evaluation}. The data extraction criteria are as follows:

\begin{itemize}
    \item \textcolor{black}{\textbf{For \textit{Architecture Problem} (D1)}, we extract the context related to architecture problems from the SO posts, GitHub issues and discussions as comprehensively as possible.}
    \item  \textbf{For \textit{Architecture Decision} (D2)}, we define different criteria for each data source. For SO, we extract architecture decisions accepted by creators of the posts or received upvotes from the SO community. When multiple architecture decisions satisfy this criterion, we prioritize those accepted by creators; if none are accepted, we choose the one with the highest number of upvotes. For GitHub issues, we extract architecture decisions that have been agreed upon in the issue conversations.  For GitHub discussions, we choose architecture decisions that have been marked as the ``Answer''.
    \item \textbf{For \textit{Design Rationale} (D3)}, we extract the information of the DR related to architecture decisions as comprehensively as possible, including advantages, disadvantages, and relevant trade-offs of architecture decisions. This information forms each argument that constitutes the ``Design Rationale''.
\end{itemize}

The first and fourth authors randomly selected 5 SO posts, 5 GitHub issues, and 5 GitHub discussions to conducted a pilot data extraction. In case of any discrepancies, the second and third authors were involved in reaching a consensus. The results show that the three data items in Table~\ref{tab:dataitem} can be extracted from our dataset. 
Subsequently, the first author conducted the formal data extraction from the labeled dataset. Each extracted data item was reviewed multiple times by the first, second, third, and fourth authors to ensure accuracy. Any inconsistencies between the four authors were discussed according to the criteria of data extraction to achieve a consensus. The data extraction results are recorded in our dataset \cite{dataset}.

\begin{table}[htbp]
\caption{Extracted data items and their descriptions}
    \small
\label{tab:dataitem}
\begin{tabular}{m{0.30cm}<{\centering}m{1.8cm}m{3.9cm}m{6.5cm}}
\toprule
\textbf{\#} & \textbf{Data Item}    & \textbf{Description} &\textbf{Example (\href{https://web.archive.org/web/20250623051536/https://github.com/openziti/ziti/issues/1256}{Issue \#1256})} \\ \hline
D1          & Architecture Problem \textcolor{black}{(\(\bm{P}\))}   & The key point(s) of the architecture problem in SO posts, GitHub issues, or GitHub discussions  & \textit{If you wish to access Console, Router, or Controller via the internet, you must open ports to the public internet and assign a public DNS record. This opens Ziti services to DDoS.}         \\ \hline
D2          & Architecture Decision \textcolor{black}{(\(\bm{D}\))}  & The key point(s) of the architecture decision in SO posts, GitHub issues, or GitHub discussions &     \textit{I propose implementing UDP hole-punching on the Controller and Router and Tunneler so they can see each other without opening the Controller and Router ports to the Internet. And since the Tunneler can now see the Controller and Router, you can also generate JWT locally on the Tunneller/Client.}            \\ \hline
D3          & Design 
Rationale \textcolor{black}{(\(\bm{DR}\))} & The key point(s) of design rationale that consists of the arguments for the architecture decision in SO posts, GitHub issues, or GitHub discussions &  
\textit{UDP hole-punching in products such as ZeroTier and Nebula that may be useful for OpenZiti:}

\textbf{Argument\_1:} \textit{The supernode is a separate stand-alone service that does not influence sensitive services, such as the controller/router that does the main work and handles and enforces zero-trust, routing, and handle management. So, if the supernode is hacked, the other components will not be compromised.}

\textbf{Argument\_2:} \textit{There are many supernodes across the Internet, so it is hard to DDoS all of them simultaneously.}

\textbf{Argument\_3:} \textit{Even if all supernodes are DDoSed and down, the peers that already established connection do not care about the supernode and continue working without an issue.}

\textbf{Argument\_4:} \textit{Only one port is opened, and you must be concerned about being open on the public Internet to one service that is not viral and acceptable if not working for hours or days.}       \\ \bottomrule
\end{tabular}
\end{table}

\subsection{Design Rationale Generation}
\label{subsec:rationalegeneration}
To answer \textcolor{black}{RQ1--RQ3} (see Section~\ref{subsec:RQs}), we designed prompts for LLMs and LLM-based agents to generate DR based on the data items defined in Table~\ref{tab:dataitem}. The DR generated with three prompting strategies is recorded in the dataset \cite{dataset}, which are further discussed in Section~\ref{sec:results}.
\subsubsection{Zero-Shot (RQ1)}
Zero-shot refers to the capability of a model to understand and complete a task without being specifically trained for it, relying on its generalizability and existing knowledge. Therefore, we only provided ``Architecture Problem'' and its corresponding ``Architecture Decision'' in prompts without any example for LLMs to generate DR. To formulate effective templates for the prompts, we conducted a pilot experiment on 10 randomly selected samples from our dataset. During the pilot experiment, we tried several templates. For example, in one template, we added extra information asking LLMs to act as an ``architecture decision expert'', detailing their specialties and responsibilities. In another template, we used a shorter instruction simply requesting LLMs to generate DR. By manually experimenting with different prompts on a subset of samples, we identified the most effective template. As shown in Fig.~\ref{fig:zeroshotprompt}, this template helps LLMs understand architecture problems and generate desirable outputs, and its effectiveness was further validated in formal experiments.

\begin{figure}[h]
    \begin{tcolorbox}[sharp corners, width=\textwidth]
    \textbf{\textit{Prompt (zero-shot):}}\\
        I will provide an architecture-related problem and the corresponding architecture decision.\\
        \textbf{Architecture Problem:} \{\textit{architecture\_problem\}}\\
        \textbf{Architecture Decision:} \{\textit{architecture\_decision\}}\\
        Please provide the design rationale of the architecture decision in light of the architecture problem.
    \end{tcolorbox}
    \caption{A template of the zero-shot method to generate DR}
    \label{fig:zeroshotprompt}
\end{figure}

\subsubsection{CoT (RQ2)}
CoT allows LLMs to progressively demonstrate their thought process before arriving at a final answer, enhancing their reasoning ability and the transparency of their decisions \cite{jason2022chain}. Therefore, in addition to ``Architecture Problem'' and ``Architecture Decision'', we also provided LLMs with a framework for analyzing the rationale in prompts. 
We designed several versions of the framework for LLMs to analyze the rationale and conducted pilot experiments on 10 randomly selected samples from our dataset. According to the results of the pilot experiments, we identified the most efficient CoT framework with four steps shown in Fig.~\ref{fig:cotprompt}. In \textbf{Step 1}, we ask the LLMs to understand architecture decisions based on the context of architecture problems. 
% The output generated by the LLMs in Step 1 helps to determine whether they have fully understood the architecture problems and  their decisions. 
In \textbf{Step 2} and \textbf{Step 3}, we ask the LLMs to analyze the advantages and disadvantages of the architecture decisions, respectively. Finally, in \textbf{Step 4}, we ask the LLMs to present the DR based on the trade-off of the advantages and disadvantages. 

It is important to note that we did not provide any other examples in prompts for LLMs. This is because the main purpose of providing additional examples in CoT is to help the model understand the required reasoning process \cite{jason2022chain}. However, the results of the pilot experiments demonstrate that LLMs can generate DR step by step according to the designed CoT framework without any examples, which was further confirmed in the formal experiments. In addition, introducing examples in prompts \commentword{significantly} increases the length of input text, which raises the difficulty for LLMs to understand architecture problems and leads to their generated DR being potentially influenced by specific examples. 

\begin{figure}[h]
    \begin{tcolorbox}[sharp corners, width=\textwidth]
    \textbf{\textit{Prompt (CoT):}}\\
        I will provide an architecture-related problem and the corresponding architecture decision.\\
        \textbf{Architecture Problem:} \{\textit{architecture\_problem\}}\\
        \textbf{Architecture Decision:} \{\textit{architecture\_decision\}}\\
        Please provide the design rationale of the architecture decision in light of the architecture problem.
        Let's think step by step:\\
        \textbf{Step 1:} Understand the architecture decision based on the provided architecture problem.\\
        \textbf{Step 2:} Analyze the advantages of choosing this architecture decision.\\
        \textbf{Step 3:} Analyze the disadvantages of choosing this architecture decision.\\
        \textbf{Step 4:} Considering the advantages and disadvantages with the trade-offs between them, please provide a detailed design rationale for choosing this architecture decision.
    \end{tcolorbox}
    \caption{A template of the CoT method to generate DR}
    \label{fig:cotprompt}
\end{figure}

\subsubsection{LLM-based Agents (RQ3)}
To answer RQ3 in Section~\ref{subsec:RQs}, we designed five LLM-based agents (i.e., \textit{Aspect\_Identifier}, \textit{Information\_Collector}, \textit{Aspect\_Analyst}, \textit{Aspect\_Reviewer}, and \textit{Trade-off\_Analyst}) by utilizing the MetaGPT framework \cite{hong2023metagpt}. MetaGPT utilizes an assembly line approach to assign different roles to various agents, effectively breaking down complex tasks into subtasks that require collaboration among multiple agents. According to the study of Liu \textit{et al.} \cite{liu2024agent}, the design pattern of MetaGPT is \textbf{role-based cooperation}, which has been employed in many SE studies (e.g., \cite{jin2024mare, widyasari2024codeagent}). Currently, no research has reached a widespread consensus on the role delineation for generating DR. Thus, we approached a process of generating DR from the perspective of architects \cite{jansen2008documenting, arman2018recovering}, breaking it down into a set of subtasks assigned to five distinct agents. 

\begin{figure}[h]
    \centering
    \includegraphics[width=\textwidth]{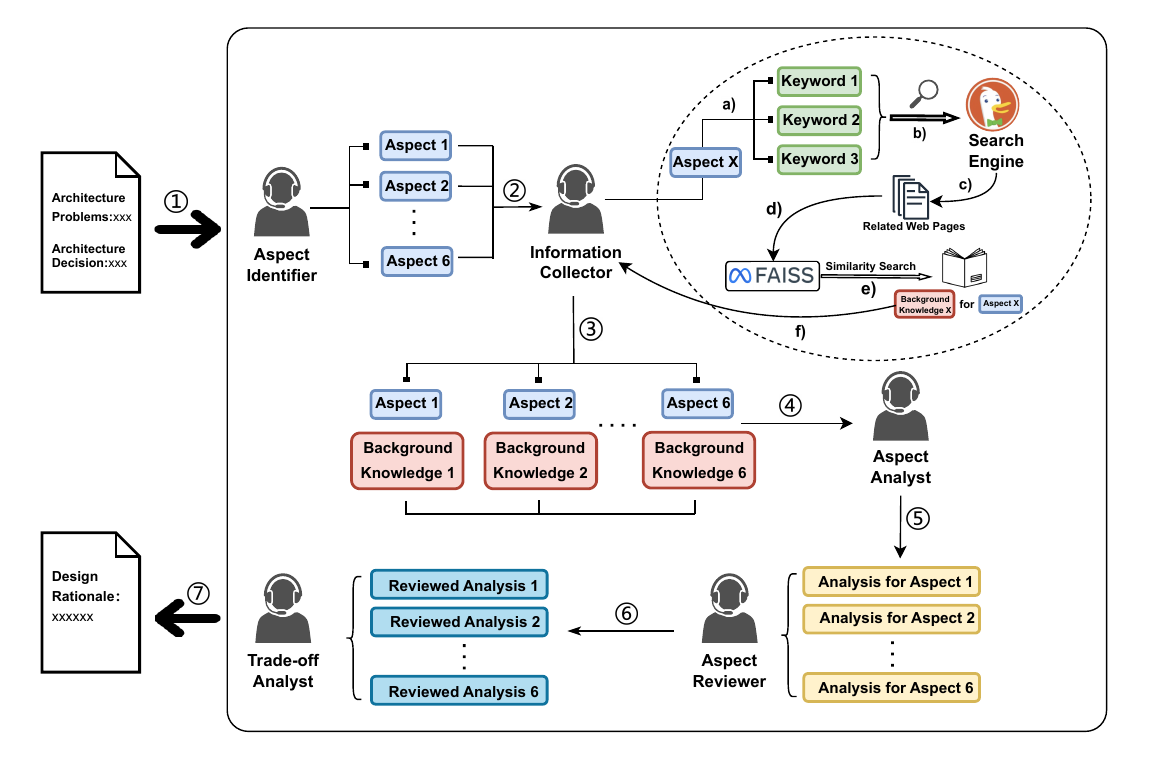}
    \begin{minipage}{\textwidth} 
    \vspace{0.1cm}
    \vspace{0.1cm}
    \scriptsize  \textbf{Note:} Steps \textcircled{1}-\textcircled{7} represent the key steps in the DR generation process based on LLM-based agents. Steps \textbf{a)} - \textbf{f)} represent the specific procedures employed by \textit{Information\_Collector} agent to gather background knowledge for \textbf{Aspect X}.
    \end{minipage}
    \caption{Overview of collaboration among LLM-based agents to generate DR}
    \label{fig:llmbasedagents}
\end{figure}

The process for generating DR using LLM-based agents is the following: (1) The architecture problems and the corresponding architecture decisions are provided to the agent system. (2) \textit{Aspect\_Identifier} agent identifies no more than six relevant aspects for analysis. (3) \textit{Information\_Collector} agent gathers relevant background information for each aspect. (4) The aspects and their background knowledge are provided to \textit{Aspect\_Analyst} agent for analysis. (5) The analysis results from \textit{Aspect\_Analyst} agent are reviewed and modified by \textit{Aspect\_Reviewer} agent. (6) \textit{Trade-off\_Analyst} agent compiles all analysis results of the aspects and generates the final DR. (7) Eventually, the final DR is provided as the output of the agent system. The overview of collaboration between the five agents is presented in Fig.~\ref{fig:llmbasedagents}.

The introduction to the five LLM-based agents is as follows:
\begin{itemize}
    \item \textbf{\textit{Aspect\_Identifier}} agent is responsible for identifying the most relevant and important aspects related to certain architecture decisions. Specifically, \textit{Aspect\_Identifier} agent receives the architecture problem and the corresponding architecture decision, and then provides up to six of the most relevant aspects, along with their definitions. We set this upper limit at six aspects according to the results of the pilot experiments on 10 samples from our dataset. The results indicated that without such a restriction, LLMs tend to enumerate too many aspects, which can make it difficult to maintain a clear focus in the subsequent analysis of DR. Additionally, this excessive enumeration \commentword{significantly} increases the costs for LLM-based agents. As a result, we found that limiting the aspects to six is considered a reasonable compromise, ensuring both depth of analysis and a concentrated focus. Then, each defined aspect is transmitted to \textit{Information\_Collector} agent.
    
    \item \textcolor{black}{\textbf{\textit{Information\_Collector}} agent is in charge of gathering relevant information through search engines.} We initially intended to enhance the information retrieval capabilities of \textit{Information\_Collector} agent by using Retrieval-Augmented Generation (RAG) \cite{gao2024retrieval} technique with a unified database. However, the diversity of architecture problems prevented us from providing a single database that could adequately address all issues. Consequently, we chose to dynamically retrieve relevant knowledge from search engines based on specific problems. Specifically, \textit{Information\_Collector} agent receives architecture problems, the corresponding architecture decisions, and aspects predefined by \textit{Aspect\_Identifier} agent. Based on each aspect, \textit{Information\_Collector} generates the three most relevant keywords for search engine queries. We chose DuckDuckGo as our search engine as it offers search results without personalized interference and provides a high level of information reliability \cite{hannak2013measuring}. For each keyword, \textit{Information\_Collector} agent retrieves the top 20 URLs returned by the search engine. Therefore, for each aspect, a total of 60 relevant URLs are collected (i.e., 3 $\times$ 20 URLs). 
    % Note that, we restrict these URLs to ensure that they are not included in our dataset, thus preventing the agent from using the original SO posts, GitHub issues, or GitHub discussions as reference information (i.e., data leakage issues). 
    To prevent the agent from using the original SO posts, GitHub issues, or GitHub discussions as reference information (i.e., data leakage issues), we discard any collected background knowledge whose URLs appear in the dataset and replace them with alternative sources.
    Then, the agent stores the webpage content from these URLs in Facebook AI Similarity Search (FAISS), a library designed for fast nearest-neighbor search and dense vector clustering \cite{douze2024faiss}. By using the similarity search function of FAISS, \textit{Information\_Collector} agent selects the 10 most relevant pieces of information from all the collected webpages that are most similar to the given architecture problems and architecture decisions. The 10 pieces of information are then transmitted to \textit{Aspect\_Analyst} agent as background knowledge along with the aspects and their definitions identified by \textit{Aspect\_Identifier} agent. This collected information has not undergone rigorous manual checking, and as a result, it may contain irrelevant or inaccurate details. We addressed this issue through \textit{Aspect\_Analyst} and \textit{Aspect\_Reviewer} agents.
    
    \item \textbf{\textit{Aspect\_Analyst}} agent is tasked with analyzing different aspects of architecture decisions. Specifically, \textit{Aspect\_Analyst} agent inputs are architecture problems, the corresponding architecture decisions, aspects that need to be analyzed, and the background knowledge collected by \textit{Information\_Collector} agent. Then, \textit{Aspect\_Analyst} agent provides a detailed analysis of the chosen architecture decisions in each aspect. Each analysis for different aspects is transmitted to \textit{Aspect\_Reviewer} agent. Additionally, since the background knowledge may contain inaccurate information, we alerted \textit{Aspect\_Analyst} agent to this risk and instructed it to verify and review the information before using it.
    
    \item \textbf{\textit{Aspect\_Reviewer}} agent is accountable for reviewing the analysis results of different aspects from \textit{Aspect\_Analyst} agent. Specifically, \textit{Aspect\_Reviewer} agent receives architecture problems, the corresponding architecture decisions, and relevant aspects along with their analysis. Note that, we did not provide the background knowledge to \textit{Aspect\_Reviewer} agent. This decision is based on our pilot experiments, which indicate that introducing background knowledge could expose \textit{Aspect\_Reviewer} agent to potential biases, hindering the ability of LLMs to maintain an independent and objective evaluation. If \textit{Aspect\_Reviewer} agent considers the analysis to be reasonable, the analysis results will be accepted. Otherwise, \textit{Aspect\_Reviewer} agent would identify the deficiencies in the analysis results and propose necessary modifications to the analysis results. The analyses that either pass the review or are modified accordingly will be forwarded to \textit{Trade-off Analyst} agent for final evaluation.
    
    \item \textbf{\textit{Trade-off\_Analyst}} agent needs to provide a final DR for the chosen architecture decisions based on all previous analysis results. Specifically, \textit{Trade-off\_Analyst} agent receives architecture problems, the corresponding architecture decisions, and the reviewed analysis results provided by \textit{Aspect\_Reviewer} agent. Then, \textit{Trade-off\_Analyst} agent analyzes the trade-offs to evaluate the architecture decisions and provides a final DR explaining why this specific design decision was made.
\end{itemize}

Liu \textit{et al.} \cite{liu2024agent} identified that the main drawback of role-based collaboration in multi-agent design involves increased communication overhead due to collaboration between agents, and the possibility of varying prices for agent services with different roles. In our pilot experiments, we found that making the interaction process of agents overly complex did not greatly improve the quality of the generated DR. Therefore, we referenced the \textbf{single-path plan generator} design pattern \cite{liu2024agent}, making each agent have only one subsequent step after completing its task, which is to submit the necessary information from the task execution results to the next agent. 

\subsection{Evaluation Methods}\label{subsec:evaluation}
\textcolor{black}{According to Equation~\ref{eq:DRdefinition}, each DR consists of multiple arguments \(\{\bm{A}_1, \bm{A}_2, \ldots, \bm{A}_n\}\). Using individual arguments in DR as the fundamental unit of analysis, we applied both quantitative accuracy metrics and the qualitative IHUM-category to comprehensively evaluate the generated DR.} Accuracy metrics, including Precision, Recall, and F1-score, are employed for quantitative analysis. IHUM-category, a classification we defined in this study, is used for the qualitative analysis of DR, which is detailed in Section~\ref{subsubsec:IHUM}. The pilot evaluation was conducted by the first and fourth authors. They randomly selected 15 DR generated by LLMs to conduct a pilot evaluation: five generated with zero-shot, five with CoT, and five with LLM-based agents. The inter-rater reliability between the two authors was measured by the Cohen's Kappa coefficient \citep{jacob1960coefficient}. The Cohen's Kappa value for the accuracy metrics is 0.863, whereas for IHUM-category is 0.738, indicating almost perfect and substantial agreement between the two authors, respectively. Then, the second and third authors joined the discussion with the first and fourth authors to resolve discrepancies and reach a consensus on the evaluation criteria. Subsequently, the first author conducted a formal evaluation of all generated DR. The evaluation results were thoroughly reviewed by the four authors (the first to fourth authors), with discrepancies resolved through discussion to reach a consensus. It is important to note that we did not use text similarity and semantic similarity metrics such as BLEU \cite{papineni2002bleu}, ROUGE \cite{lin2004rouge}, and METEOR \cite{banerjee2005meteor}, to evaluate the LLM-generated DR. Based on the pilot experiments conducted on 10 randomly selected DR generated by LLMs, we observed a significant difference between the text provided by human experts and that generated by LLMs. Simply applying text and semantic similarity metrics yields very low values with all three prompting strategies, making it challenging to draw valid experimental conclusions.

\subsubsection{Accuracy Metrics}
For accuracy metrics, we used Precision, Recall, and F1-score to evaluate the alignment of DR generated by LLMs with those provided by human experts. The three metrics are commonly used to measure the quality of DR (e.g., \cite{zhao2024drminer, soliman2025do}). As mentioned in Section~\ref{sec:background}, DR comprises a set of arguments. We compared the arguments provided in the DR generated by LLMs with those extracted from human experts during the data extraction phase. We then labeled the arguments into true positive (TP), false positive (FP), and false negative (FN). To be specific, we set the following criteria for labeling arguments:

\begin{itemize}
    \item \textbf{TP:} The argument points present the DR provided by both LLMs and human experts.
    \item \textbf{FP:} The argument points that are present exclusively in the DR generated by LLMs and do not align with the ground truth provided by human experts.
    \item \textbf{FN:} The argument points that are missing from the DR generated by LLMs but are present in the ground truth provided by human experts.
    \item \textbf{TN:} The argument points that are absent from both the LLM-generated DR and the ground truth provided by human experts. Such argument points are not included in our evaluation, as they do not exist.
\end{itemize}

Equations ~\eqref{eq:precision}, ~\eqref{eq:recall} and ~\eqref{eq:F1-score} present three accuracy metrics, Precision, Recall, and F1-score. For a given data point \((P_i, D_i, DR_i)\), \(TP_i\), \(FP_i\), and \(FN_i\) are the number of arguments classified as true positive, false positive, and false negative in the DR. As the number of arguments from human experts is typically small (usually less than three), slight variations in the classification of arguments as \(TP\), \(FP\), and \(FN\) can cause significant fluctuations in the metrics. For instance, when two arguments are provided by an expert, an increase in \(TP_i\) from one to two results in a sharp rise in Recall from 50\% to 100\%. \textcolor{black}{Therefore, these accuracy metrics represent a weighted average over the 100 data points, with weights determined by the number of arguments provided in each DR. For example, weighted Precision is calculated by averaging the per-DR precision scores, weighted by the number of predicted arguments (i.e., \(TP_i\)+\(FP_i\)) in each DR.}

\begin{equation}
\label{eq:precision}
\text{Precision} = \textcolor{black}{\sum \left( \frac{TP_i + FP_i}{\sum (TP_j + FP_j)} \frac{TP_i}{TP_i + FP_i}  \right )} = \frac{\sum (TP_i)}{\sum (TP_i + FP_i)}
\end{equation}

\begin{equation}
\label{eq:recall}
\text{Recall} = \textcolor{black}{\sum \left( \frac{TP_i + FN_i}{\sum (TP_j + FN_j)} \frac{TP_i}{TP_i + FN_i} \right)} = \frac{\sum (TP_i)}{\sum (TP_i + FN_i)}
\end{equation}

\begin{equation}
\label{eq:F1-score}
\text{F1-score} = \frac{2 \times \text{Precision} \times \text{Recall}}{\text{Precision} + \text{Recall}}
\end{equation}

\subsubsection{IHUM-Category}\label{subsubsec:IHUM}
We introduced the IHUM-category to qualitatively assess the argument points in DR generated by LLMs. Accuracy metrics are used to compare the similarity between the DR generated by LLMs and that provided by human experts. However, there may be a chance that the responses generated by LLMs do not align with those of human experts (i.e., FN); the arguments still offer valuable insights and provide alternative perspectives. We proposed that the evaluation of LLM-generated DR could follow a methodology akin to the one employed by Mahajan \textit{et al.} \cite{mahajan2020recommending} for assessing responses on SO. Therefore, we introduced the IHUM-category, which is adapted from the metrics in \cite{mahajan2020recommending}, with adjustments made to effectively assess DR. IHUM-category is used to manually classify all the argument points, allowing for a more comprehensive evaluation of the DR generated by LLMs. The definition of IHUM-category is presented in Table~\ref{tab:IHUMcategory}.

\begin{table}[htbp]
\caption{IHUM-category for classifying argument points in LLM-generated DR}
\small
\label{tab:IHUMcategory}
\begin{tabular}{m{2cm}m{11cm}}
\hline
\textbf{Label}     & \textbf{Definition}                     \\ \hline
Insightful (I)     & The argument provides correct and valuable perspectives that are most relevant to the architecture decision. Given the difficulty of assessing the relevance of an argument to the architecture decision, we refer to the rationale provided by human experts. Specifically, if the argument generated by LLMs aligns with the argument points raised by human experts, we consider it to be ``Insightful''.  \\ \hline
Helpful (H)        & The argument provides a correct perspective that contributes to the architecture decision, but may not be the most critical reason for making this design decision.  \\ \hline
Uncertain (U)      & The argument is difficult to assess for correctness due to vague descriptions or insufficient contextual information.   \\ \hline
Misleading (M)     & The argument provides an incorrect or misleading perspective.   \\ \hline
\end{tabular}
\end{table}

\textcolor{black}{As part of our evaluation, we calculated the proportion of each of the four IHUM categories and used them as assessment metrics. Equation~\ref{eq:IHUM} illustrates the computation of the proportion of ``Insightful'' arguments, with analogous formulas applied to the other categories (``Helpful'', ``Uncertain'', and ``Misleading''). To maintain consistency with the accuracy metrics, a weighted calculation based on the number of arguments is employed.}

\begin{equation}
\label{eq:IHUM}
\textcolor{black}{\frac{I}{I+H+U+M} = \sum \left( \frac{I_i+H_i+U_i+M_i}{\sum (I_j+H_j+U_j+M_j)} \frac{I_i}{I_i+H_i+U_i+M_i}  \right ) = \frac{\sum (I_i)}{\sum (I_i+H_i+U_i+M_i)}}
\end{equation}

\subsection{Interviews with Practitioners}
\label{subsec:interviews_steps}
\textcolor{black}{To answer RQ4, we conducted semi-structured interviews to understand the practitioners' perceptions of the LLM-generated DR by following the guidelines proposed by Wohlin \textit{et al.} \cite{Wohlin2012ESE}. Semi-structured interviews enable practitioners to talk deeply and freely about their perspectives on the DR generated by LLMs based on their experience. The interview consists of three steps:}
\begin{itemize}
    \item \textcolor{black}{\textbf{Step 1}: Before the interviews, participants were introduced to the research objectives and methodology of our study. They were then asked to review five randomly selected samples from our dataset \cite{dataset}, comprising architecture problems and decisions, along with the corresponding DR provided by human experts and generated by LLMs. The five samples were compiled in a PDF file and distributed to the participants for review prior to the interviews. The PDF file that contains the five samples has been provided in our dataset \cite{dataset}.}
    \item \textcolor{black}{\textbf{Step 2}: The interviews were conducted by the first author. Each participant provided a brief overview of their background, including their professional role, as well as their years of experience in software development and architecture design.}
    \item \textcolor{black}{\textbf{Step 3}: During the semi-structured interviews, practitioners responded to five Interview Questions (IQs). With participants' consent, the interviews were audio-recorded to facilitate subsequent analysis.}
\end{itemize}

\textcolor{black}{The five IQs were designed to explore two key aspects of LLM-generated DR from the perspective of practitioners: \textbf{trustworthiness} and \textbf{applicability}. Specifically, IQ1.1 and IQ1.2 focus on participants' level of trust in the DR produced by LLMs and the types of information that could enhance their trust. IQ2.1 to IQ2.3 examine the potential use of LLM-generated DR in practice, including possible application scenarios and perceived challenges. The five IQs are as follows:}
\begin{itemize}
    \item \textcolor{black}{\textbf{IQ1.1}: To what extent would you trust the DR generated by LLMs for architecture decisions? Why? (Options range from ``Do not trust at all'' to ``Fully trust without human validation''.)}
    \item \textcolor{black}{\textbf{IQ1.2}: What types of information (e.g., project context, domain-specific knowledge) should be provided to LLMs to enhance the trustworthiness of the generated DR for architecture decisions?}
    \item \textcolor{black}{\textbf{IQ2.1}: Would you consider using LLM-generated DR in practice? (Options range from ``Definitely not'' to ``Definitely yes''.)}
    \item \textcolor{black}{\textbf{IQ2.2}: What scenarios (e.g., comparing architectural alternatives, understanding architecture design) might be suitable for applying LLM-generated DR for architecture decisions?}
    \item \textcolor{black}{\textbf{IQ2.3}: What are the main challenges or limitations in applying LLM-generated DR in practice?}
\end{itemize}

\section{Results and Interpretation}\label{sec:results}
In this section, we present the results of the \textcolor{black}{four} RQs with their interpretation. Specifically, we provide the accuracy metrics and IHUM-category classification results of the DR generated by the five LLMs (see Section~\ref{subsec:LLMselection}) using three distinct prompting strategies (see Section~\ref{subsec:rationalegeneration}), detailed in Tables~\ref{tab:resultsofRQ1},~\ref{tab:resultsofRQ2}, and~\ref{tab:resultsofRQ3}, respectively. \textcolor{black}{Table~\ref{tab:resultsofRQ1} to Table~\ref{tab:resultsofRQ3} additionally report the average performance of the metrics for DR generation, highlighting how LLMs perform under each prompting strategy and supports the interpretation of results in Section~\ref{sec:results} as well as the discussion in Section~\ref{sec:discussion}.} Table~\ref{tab:comparison} presents the average performance of the five LLMs in generating DR with three prompting strategies.
% It should be noted that, the proportion of arguments classified under the ``Insight'' category in the DR is essentially equivalent to ``Precision'' in the accuracy metrics, as defined in Section~\ref{sec:methodology}.
\textcolor{black}{In Table~\ref{tab:resultsofRQ1} to Table~\ref{tab:comparison}, the best-performing results of the five LLMs are highlighted using bold and underlined text. Notably, since the number and proportion of arguments in the ``Uncertain'' and ``Misleading'' categories are negative indicators, the smallest values are highlighted in the tables.}

\subsection{RQ1: How effective are LLMs in generating and recovering design rationale for architecture decisions?}
Table~\ref{tab:resultsofRQ1} presents the accuracy metrics (including Precision, Recall, and F1-score) and IHUM-category classification results of the DR generated by the five LLMs with the zero-shot method.

\begin{table}[htbp]
\vspace{0.5cm}
\caption{The evaluation results of DR generated with the zero-shot method}
\vspace{0.2cm}
\label{tab:resultsofRQ1}
\centering
\footnotesize
\begin{tabular}{m{2.3cm}ccccc|c}
\hline
\rowcolor{lightgray}  & \textbf{gpt-3.5-turbo} & \textbf{gpt-4-0613} & \textbf{gemini-1.0-pro} & \textbf{llama3-8B} & \textbf{mistral-7B} &  \textbf{Average} \\ \hline
\textbf{Precision} & 0.265 & 0.250 & \textbf{\underline{0.295}} & 0.286 & 0.277 & 0.278 \\ 
\textbf{Recall} & 0.591 & \textbf{\underline{0.697}} & 0.652 & 0.576 & 0.621 & 0.627 \\ 
\textbf{F1-score} & 0.366 & 0.368 & \textbf{\underline{0.406}} & 0.382 & 0.383 & 0.381 \\ 
\textbf{Argument count} & 441 & \textbf{\underline{553}} & 438 & 399 & 444 & 455 \\ \hline
\textbf{Insightful (I)}   & 117 & \textbf{\underline{138}} & 129  & 114 & 123 & 124.2 \\ 
\textbf{Helpful (H)} & 290 & \textbf{\underline{384}} & 273 & 246 & 280 & 294.6 \\ 
\textbf{Uncertain (U)} & 22 & 23 & 22 & \textbf{\underline{19}} & 24 & 22 \\ 
\textbf{Misleading (M)} & 12 & \textbf{\underline{8}} & 14 & 20 & 17 & 14.2 \\ 
\textbf{I/(I+H+U+M)} & 26.53\% & 24.96\% & \textbf{\underline{29.45\%}} & 28.57\% & 27.70\% & 27.44\%\\ 
\textbf{H/(I+H+U+M)} & 65.76\% & \textbf{\underline{69.44\%}} & 62.33\% & 61.65\% & 63.06\% & 64.45\%\\ 
\textbf{U/(I+H+U+M)} & 4.99\% & \textbf{\underline{4.16}}\% & 5.02\% & 4.76\% & 5.41\% & 4.87\%  \\ 
\textbf{M/(I+H+U+M)} & 2.72\% & \textbf{\underline{1.45\%}} & 3.20\% & 5.01\% & 3.83\% & 3.24\%  \\ \hline
\end{tabular}
\vspace{0.2cm}\par
\begin{minipage}{\linewidth}
  \hspace*{1em}
  \footnotesize
  \textcolor{black}{\textbf{Note:} Best results among the five LLMs are highlighted using underlining and bold font, such as \underline{\textbf{xx.xx\%}}}.
\end{minipage}
\vspace{0.2cm}
\end{table}

\subsubsection{For Accuracy Metrics}
The average Precision, Recall, and F1-score of the DR generated by the five LLMs with the zero-shot method are 0.278, 0.627, and 0.381, respectively. Among the five LLMs, gemini-1.0-pro exhibits the highest Precision (0.295) and F1-score (0.406), while gpt-4-0613 achieves the highest Recall (0.697).

\textcolor{black}{The results of the zero-shot method show the capability of LLMs in generating DR using a basic prompting strategy. According to the accuracy results, about one in four arguments generated by the LLMs are consistent with those from human experts. Additionally, over 60\% of the expert-provided arguments are also mentioned in the LLM-generated DR. The five LLMs exhibited relatively low precision, which in turn contributed to the low F1-score. However, this is primarily due to the tendency of LLMs to generate a larger number of arguments. In fact, the average number of arguments generated by the LLMs with zero-shot (4.55) is 2.3 times greater than that provided by human experts (1.98). Therefore, even though the LLMs' generated DR covers most of the arguments provided by human experts, the Precision still does not exceed 0.3. Similar patterns are observed in both the CoT and LLM-based agent methods. The arguments generated by LLMs but not considered by human experts are further evaluated using the IHUM-category, as presented in Section~\ref{subsubsec:zero-shot_IHUM}. However, an issue of zero-shot method lies in that the arguments generated by LLMs tend to emphasize the advantages of the chosen architecture while neglecting its potential disadvantages. As a consequence, when human experts focus on architectural shortcomings in some architecture problems (e.g., \href{https://web.archive.org/web/20250623053253/https://github.com/Expensify/App/issues/33070}{Issue \#33070} and \href{https://web.archive.org/web/20250623042214/https://github.com/dotnet/wpf/discussions/9022}{Discussion \#214}), the Precision of the DR produced by LLMs under the zero-shot method is low. The limitation due to the absence of architectural shortcomings of the zero-shot method is further discussed in Section~\ref{subsec:pros&cons}.}
\vspace{0.2cm}

\label{subsubsec:zero-shot_IHUM}
\subsubsection{For IHUM-Category} The average proportion of arguments classified under ``Insightful'', ``Helpful'', ``Uncertain'', and ``Misleading'' categories are 27.44\%, 64.45\%, 4.87\%, and 3.24\%, respectively. It is worth noting that besides the 27.44\% arguments that are consistent with human experts (``Insightful''), around 65\% are not considered by the human experts, but also offer ``Helpful'' insights into architecture design. Additionally, 4.87\% of the arguments are indeterminate (``Uncertain'') and 3.24\% contain potential errors (``Misleading''), representing a relatively small proportion. Among the five LLMs, gemini-1.0-pro generated the highest proportion of ``Insightful'' arguments (29.45\%), while gpt-4-0613 produces the highest proportion of ``Helpful'' arguments (69.44\%), along with the lowest proportions in the ``Uncertain'' (4.16\%) and ``Misleading'' categories (1.45\%). 

\textcolor{black}{Among the three prompting strategies, the DR generated by the zero-shot method exhibits the highest proportion of ``Misleading'' arguments. This result may be attributed to the prompt template of zero-shot, which directly instructs the LLMs to generate DR without incorporating an explicit reasoning process. The absence of such guided reasoning contributes to the lack of trade-off analysis in the DR. Besides, in addition to the arguments that can be categorized using the IHUM-category, the DR generated by the zero-shot method occasionally contains content that is unrelated to the rationale of architecture decisions (e.g.,  alternative architectures, code snippets). This tendency is especially evident in the DR generated by gpt-4-0613 and llama3-8B, which contain a higher proportion of such extraneous information than other LLMs. The limitations of zero-shot are further discussed in Section~\ref{subsec:pros&cons}.}

\vspace{0.6cm}
\begin{tcolorbox}[colback=gray!20, colframe=gray]
\textbf{Key findings of RQ1}
\begin{itemize}
    \item With the zero-shot method, the five LLMs are capable of generating DR for the architecture decisions in our dataset.
    \item About 28\% of the arguments in the DR generated by LLMs align with the insights from human experts, representing 62\% of the total arguments the human experts provided.
    \item The higher number of arguments in the DR generated by the LLMs compared to those provided by human experts is the reason for its relatively low Precision and F1-score.
    \item About 65\% of the arguments in the generated DR provide ``Helpful'' insights that were not considered by human experts. Besides, fewer than 5\% of the arguments are ``Uncertain'' in correctness, while 3.24\% are ``Misleading''.
\end{itemize}
\end{tcolorbox}
\vspace{0.8cm}

\subsection{RQ2: Does the CoT technique affect or enhance the capability of LLMs to generate and recover design rationale for architecture decisions?}

Table~\ref{tab:resultsofRQ2} presents the accuracy metrics (including Precision, Recall, and F1-score) and IHUM-category classification results of the DR generated by the five LLMs with the CoT method.

\begin{table}[h!]
\vspace{0.5cm}
\caption{The evaluation results of DR generated with the CoT method}
\label{tab:resultsofRQ2}
\centering
\footnotesize
\resizebox{\textwidth}{!}{
\begin{tabular}{m{2.3cm}ccccc|cc}
\hline
\rowcolor{lightgray}  & \textbf{gpt-3.5-turbo} & \textbf{gpt-4-0613} & \textbf{gemini-1.0-pro} & \textbf{llama3-8B} & \textbf{mistral-7B} & \textbf{Average} & \textcolor{black}{\textbf{zero-shot}} \\ \hline
\textbf{Precision}  & \textbf{\underline{0.280}} & 0.231 & 0.224 & 0.235 & 0.214 & 0.237 & \textcolor{black}{0.278} \\ 
\textbf{Recall} & 0.682 & \textbf{\underline{0.717}} & 0.657 & 0.667 & 0.697 & 0.684 & \textcolor{black}{0.627} \\ 
\textbf{F1-score} & \textbf{\underline{0.396}} & 0.350 & 0.335 & 0.347 & 0.327 & 0.351 & \textcolor{black}{0.381} \\ 
\textbf{Argument count} & 483 & 614 & 578 & 562 & \textbf{\underline{646}} & 576.6 & \textcolor{black}{455} \\ \hline
\textbf{Insightful (I)}   & 135 & \textbf{\underline{142}} & 130  & 132 & 138 & 135.4 & \textcolor{black}{124.2} \\ 
\textbf{Helpful (H)} & 314 & \textbf{\underline{442}} & 400 & 389 & 463 & 401.6 & \textcolor{black}{294.6} \\ 
\textbf{Uncertain (U)} & 25 & \textbf{\underline{23}} & 30 & \textbf{\underline{23}} & 30 & 26.2 & \textcolor{black}{22} \\ 
\textbf{Misleading (M)} & 9 & \textbf{\underline{7}} & 18 & 18 & 15 & 13.4 & \textcolor{black}{14.2} \\ 
\textbf{I/(I+H+U+M)} & \textbf{\underline{27.95\%}} & 23.13\% & 22.49\% & 23.49\% & 21.36\% & 23.68\% & \textcolor{black}{27.44\%}\\ 
\textbf{H/(I+H+U+M)} & 65.01\% & \textbf{\underline{71.99\%}} & 69.20\% & 69.22\% & 71.67\% & 69.42\% & \textcolor{black}{64.45\%}\\ 
\textbf{U/(I+H+U+M)} & 5.18\% & \textbf{\underline{3.75\%}} & 5.19\% & 4.09\% & 4.64\% & 4.57\% & \textcolor{black}{4.87\%} \\ 
\textbf{M/(I+H+U+M)} & 1.86\% & \textbf{\underline{1.14}}\% & 3.11\% & 3.20\% & 3.32\% & 2.33\% & \textcolor{black}{3.24\%} \\ \hline
\end{tabular}
}
\vspace{0.2cm}\par
\begin{minipage}{\linewidth}
  \hspace*{1em}
  \footnotesize
  \textcolor{black}{\textbf{Note:} Best performance among the five LLMs are highlighted using underlining and bold font, such as \underline{\textbf{xx.xx\%}}}.
\end{minipage}
\vspace{0.3cm}
\end{table}

\subsubsection{For Accuracy Metrics} 
The average Precision, Recall, and F1-score of the DR generated by the five LLMs with the CoT method are 0.237, 0.684, and 0.351, respectively. Among the five LLMs, gpt-3.5-turbo demonstrates the highest Precision (0.280) and F1-score (0.396), while gpt-4-0613 attains the highest Recall (0.717).

Except for gpt-3.5-turbo, the performance of the other four LLMs in terms of Precision and F1-score deteriorates with the CoT method, compared to the performance of the zero-shot method. This can be explained by the \commentword{significantly} higher number of arguments generated in the DR when using the CoT template, reaching 5.76 per DR on average, compared to 4.55 per DR with the zero-shot method. \textcolor{black}{Among the five LLMs, mistral-7B exhibits the largest increase in the number of arguments (+2.0), followed by llama3-8B (+1.6) and  gemini-1.0-Pro (+1.4), while gpt-4-0613 (+0.6) and gpt-3.5-turbo (+0.4) show relatively smaller increases.} The increased number of arguments stems from the CoT template explicitly guiding LLMs to conduct a trade-off analysis after providing both the advantages and disadvantages. In contrast, the DR generated with the zero-shot template is less likely to consider the negative impacts of architecture decisions. Due to the increase in the number of arguments generated with the CoT template, the DR considers more diverse perspectives compared to the zero-shot method, which leads to a 9.1\% increase in Recall. Additionally, the decrease in Precision and F1-score is also primarily attributed to the increase in the total number of arguments generated in the DR, which leads to a lower proportion of arguments consistent with those provided by human experts. However, this does not imply a decline in the quality of DR generated using the CoT method, as we cannot expect the LLM-generated DR to consist solely of arguments provided by human experts. Arguments not considered by human experts in the LLM-generated DR can also potentially offer valuable insights, which are further evaluated according to the IHUM-category in Section~\ref{subsubsec:CoT_manual}.

\subsubsection{For IHUM-Category}
\label{subsubsec:CoT_manual}
The average proportion of arguments classified under ``Insightful'', ``Helpful'', ``Uncertain'' and ``Misleading'' categories are 23.68\%, 69.42\%, 4.57\%, and 2.33\%, respectively. Among the five LLMs, gpt-3.5-turbo exhibits a notable increase in the proportion of ``Insightful'' arguments (27.95\%), ranking the highest. Meanwhile, gpt-4-0613 maintains the highest proportion of ``Helpful'' (71.99\%) arguments and has the lowest proportion of ``Uncertain'' (3.75\%) and ``Misleading'' arguments (1.14\%).

Table~\ref{tab:resultsofRQ2} also presents a comparison of the average performance between the zero-shot and CoT methods. Although the proportion of ``Insightful'' arguments decreases, the absolute number of ``Insightful'' arguments generated with the CoT method increases, which is also reflected in the higher average Recall. Additionally, compared to the zero-shot method, the number of ``Helpful'' arguments increases by 36.3\%, while the proportion increases by 7.7\%. This improvement indicates that, with the CoT method, LLMs not only generate more ``Insightful'' arguments but also present a broader range of ``Helpful'' arguments, offering architects a more comprehensive perspective on architecture decisions. However, in certain DR, such as the one generated by gpt-4-0613 for \href{https://web.archive.org/web/20250622091847/https://stackoverflow.com/questions/78536441/separate-module-for-activity-broadcast-receiver-and-services}{SO \#78536441} and mistral-7B for \href{https://web.archive.org/web/20250623042214/https://github.com/dotnet/wpf/discussions/9022}{Discussion \#9022}, we identified an excess of arguments, which are 14 and 10 respectively. Some arguments that are less relevant to the architecture decision may result in the DR appearing overly lengthy and unfocused. Potential issues of the CoT method are further discussed in Section~\ref{subsec:pros&cons}.

The proportion of ``Uncertain'' arguments has not \commentword{significantly} decreased. However, the proportion of ``Misleading'' arguments in the DR has declined. It is noticed that the CoT method enhances the reasoning process by outlining the strengths and drawbacks of architecture decisions for conducting a trade-off analysis, thereby mitigating some of the errors observed in the zero-shot approach. For example, \href{https://web.archive.org/web/20250622085527/https://stackoverflow.com/questions/77745340/how-to-make-two-projects-with-incompatible-net-versions-talk}{SO~\#77745340} examines the rationale for \textit{``running .NET projects targeting different framework versions in separate processes''}, with Inter Process Communication (IPC) mechanisms implied. The DR generated by gpt-3.5-turbo using the zero-shot method suggests that ``\textit{running components in separate processes can utilize system resources more efficiently}'', concluding that the whole system performance is enhanced. However, the introduction of IPC actually increases the system's performance costs due to the need for serialization and deserialization during data transmission between different processes. While the DR generated with the zero-shot method ignores this issue (the cost of IPC), the DR generated by the CoT method considers the issue when analyzing the disadvantages of this architecture, ultimately leading to a more accurate trade-off conclusion. \textcolor{black}{Besides, with the CoT method, the reasoning steps are explicitly specified in the prompt template, leading all LLMs to adopt a step-by-step structure when generating DR. As a result, the DR produced by the five LLMs exhibits a high degree of consistency in content organization. Notably, gemini-1.0-pro, llama3-8B, and mistral-7B tend to include a concluding summary at the end of the DR.}

\begin{tcolorbox}[colback=gray!20, colframe=gray]
\textbf{Key findings of RQ2}
\begin{itemize}
    \item Compared to the zero-shot method, the DR generated through the CoT method includes more arguments aligned with human experts, leading to an increase in Recall. However, the higher number of arguments caused a decrease in both Precision and F1-score.
    \item Since the explicit specification in the CoT template to analyze both advantages and disadvantages of architecture decisions, the generated DR provides a more comprehensive perspective, including a greater number and proportion of ``Helpful'' arguments compared to the zero-shot method.
    \item By analyzing the pros and cons of architecture decisions and incorporating a trade-off, the CoT method avoids some errors introduced by the zero-shot method, leading to fewer ``Misleading'' arguments.
\end{itemize}
\end{tcolorbox}

\subsection{RQ3: Do LLM-based agents affect or enhance the capability of LLMs to generate and recover design rationale for architecture decisions?}
Table~\ref{tab:resultsofRQ3} presents the accuracy metrics (including Precision, Recall, and F1-score) and IHUM-category classification results of the DR generated by LLM-based agents.

\begin{table}[htbp]
\caption{The evaluation results of DR generated by LLM-based agents}
\label{tab:resultsofRQ3}
\centering
\footnotesize
\resizebox{\textwidth}{!}{
\begin{tabular}{m{2.3cm}ccccc|cc}
\hline
\rowcolor{lightgray}  & \textbf{gpt-3.5-turbo} & \textbf{gpt-4-0613} & \textbf{gemini-1.0-pro} & \textbf{llama3-8B} & \textbf{mistral-7B} &  \textbf{Average} &  \textbf{zero-shot} \\ \hline
\textbf{Precision} & \textbf{\underline{0.291}} & 0.255 & 0.269 & 0.254 & 0.266 & 0.267 & \textcolor{black}{0.278}\\ 
\textbf{Recall} & 0.717 & \textbf{\underline{0.747}} & 0.707 & 0.687 & 0.717 & 0.715 & \textcolor{black}{0.627}\\ 
\textbf{F1-score} & \textbf{\underline{0.414}} & 0.380 & 0.389 & 0.371 & 0.386 & 0.389 & \textcolor{black}{0.381}\\ 
\textbf{Argument count} & 488 & \textbf{\underline{580}} & 521 & 535 & 533 & 531.4 & \textcolor{black}{455}\\ \hline
\textbf{Insightful (I)}   & 142 & \textbf{\underline{148}} & 140  & 136 & 142 & 141.6 & \textcolor{black}{124.2}\\ 
\textbf{Helpful (H)} & 318 & \textbf{\underline{407}} & 345 & 365 & 363 & 359.6 & \textcolor{black}{294.6}\\ 
\textbf{Uncertain (U)} & 21 & \textbf{\underline{20}} & 26 & 21 & 21 & 21.8 & \textcolor{black}{22}\\ 
\textbf{Misleading (M)} & 7 & \textbf{\underline{5}} & 10 & 13 & 7 & 8.4 & \textcolor{black}{14.2}\\ 
\textbf{I/(I+H+U+M)} & \textbf{\underline{29.10\%}} & 25.52\% & 26.87\% & 25.42\% & 26.64\% & 26.71\% & \textcolor{black}{27.44\%}\\ 
\textbf{H/(I+H+U+M)} & 65.16\% & \textbf{\underline{70.17\%}} & 66.22\% & 68.22\% & 68.11\% & 67.58\% & \textcolor{black}{64.45\%}\\ 
\textbf{U/(I+H+U+M)} & 4.30\% & \textbf{\underline{3.45\%}} & 4.99\% & 3.93\% & 3.94\% & 4.12\% & \textcolor{black}{4.87\%}\\ 
\textbf{M/(I+H+U+M)} & 1.43\% & \textbf{\underline{0.86\%}} & 1.92\% & 2.43\% & 1.31\% & 1.59\% & \textcolor{black}{3.24\%}\\ \hline
\end{tabular}
}
\vspace{0.2cm}\par
\begin{minipage}{\linewidth}
  \hspace*{1em}
  \footnotesize
  \textcolor{black}{\textbf{Note:} Best performance among the five LLMs are highlighted using underlining and bold font, such as \underline{\textbf{xx.xx\%}}}.
\end{minipage}
\vspace{0.2cm}
\end{table}

\subsubsection{For Accuracy Metrics}
The average Precision, Recall, and F1-score of the DR generated by LLM-based agents are 0.267, 0.715, and 0.389, respectively. Among the five LLM-based agent systems, gpt-3.5-turbo achieves the highest Precision (0.291) and F1-score (0.414), while gpt-4-0613 demonstrates the highest Recall (0.747).

In Table~\ref{tab:comparison}, LLM-based agents generate more arguments than zero-shot but fewer than the CoT method, as \textit{Aspect\_Identifier} agent is required to specify up to six most relevant perspectives for architecture analysis. \textcolor{black}{Compared to the zero-shot method, llama3-8B shows the largest increase in the number of generated DR (+1.4), followed by mistral-7B (+0.9) and gemini-1.0-pro (+0.8), while gpt-3.5-turbo (+0.5) and gpt-4-0613 (+0.3) exhibit the smallest increases.} The DR generated by LLM-based agents has a lower average Precision than the zero-shot method but higher than the CoT method, while its F1-score is comparable to the zero-shot method and superior to the CoT method. \textcolor{black}{However, the slightly lower Precision of LLM-based agents compared to that of the zero-shot method does not necessarily indicate that the performance of LLM-based agents in generating DR is inferior. In fact, LLM-based agents generated most arguments that align with those provided by human experts, and show a notable improvement in Recall, with an increase of 19.8\% compared to the zero-shot method and 4.5\% compared to the CoT method. This result indicates that LLM-based agents can identify and focus on the key factors that align with human experts' arguments in architecture decisions. However, in comparison to the zero-shot method, LLM-based agents generated a larger number of arguments, including many that were not mentioned by human experts, which contributed to a modest 0.01 decrease in Precision. These additional arguments not mentioned by human experts did not necessarily diminish the overall quality of the DR generated by LLM-based agents. In fact, most of the arguments that contributed to the slight drop in Precision were considered ``Helpful'' for understanding the architecture decision, as detailed in Section~\ref{subsubsec:agents_manual}.}

\subsubsection{For IHUM-Category}
\label{subsubsec:agents_manual}
The average proportion of arguments classified under ``Insightful'', ``Helpful'', ``Uncertain'' and ``Misleading'' categories are 26.71\%, 67.58\%, 4.12\%, and 1.59\%, respectively. Among the five LLMs, gpt-3.5-turbo exhibits an increase in the proportion of ``Insightful'' arguments (29.10\%), ranking the highest. Meanwhile, gpt-4-0613 maintains the highest proportion of ``Helpful'' arguments (70.17\%) and has the lowest proportion of ``Uncertain'' (3.45\%) and ``Misleading'' arguments (0.86\%). 

Compared to zero-shot and CoT, LLM-based agents generate the highest number of ``Insightful'' arguments in the DR, achieving 1.42 on average per DR. Additionally, we also identified a substantial proportion of ``Helpful'' arguments, approximately 68\%, which is higher than the zero-shot method but slightly lower than the CoT method. 
% This indicates that LLM-based agents not only generate arguments that align with human experts but also offer more comprehensive analysis from different perspectives compared to the zero-shot method. 
% However, the relative decrease in ``Helpful'' arguments and the increase in ``Insightful'' arguments in the DR produced by LLM-based agents help avoid excessive supplementary points compared to CoT.

The proportion of ``Uncertain'' and ``Misleading'' arguments in the DR generated by LLM-based agents has decreased compared to other prompting strategies. The decline in ``Uncertain'' arguments is relatively marginal, while the reduction in ``Misleading'' arguments is more significant, with the proportion decreasing to 1.59\%. According to the output of each agent, the reduction in ``Misleading'' arguments is attributed to \textit{Aspect\_Reviewer} and \textit{Trade-off\_Analyst} agents. \textit{Aspect\_Reviewer} agent plays a critical role in identifying and rectifying errors within the arguments. For example, in the DR generated by gpt-3.5-turbo for \href{https://web.archive.org/web/20250623050259/https://github.com/ran-isenberg/aws-lambda-handler-cookbook/discussions/787}{Discussion \#787}, \textit{Aspect\_Reviewer} agent highlights that the performance analysis by \textit{Aspect\_Analyst} lacks sufficient rigor, emphasizing that factors such as concurrency, resource management, and scalability are also crucial in evaluating whether the architecture decision would lead to performance improvements. This analysis of \textit{Aspect\_Reviewer} agent effectively prevents the inclusion of vague and inadequately substantiated conclusions of ``\textit{enhancing system efficiency}'' as a reference in the final trade-off analysis. Therefore, in the final DR, the performance discussion focuses on potential concurrency issues in a multi-threaded environment and the need for effective resource management, rather than claiming a definitive performance improvement. 
\textit{Trade-off\_Analyst} agent is required to conduct a trade-off analysis according to the analysis from different aspects and provide the final DR. We observed that the DR generated by the multi-agent system based on five LLMs incorporates a consideration of both positive and negative impacts, which is similar to the DR generated with the CoT template. \textcolor{black}{Furthermore, due to the absence of structural constraints in the \textit{Trade-off\_Analyst} agent, the generated DR exhibits variation in content organization across different models. Notably, gpt-3.5-turbo and gpt-4-0613, tend to perform trade-off analysis within each argument, directly balancing the pros and cons of a specific architectural aspect. In contrast, gemini-1.0-pro, mistral-7B, and llama3-8b tend to list separate arguments, usually outlining advantages and disadvantages, and then perform a summary-level trade-off across them.}

\begin{tcolorbox}[colback=gray!20, colframe=gray]
\textbf{Key findings of RQ3}
\begin{itemize}
    \item Compared to the zero-shot and CoT methods, the DR generated by LLM-based agents includes the most arguments that align with human experts, exhibiting the highest Recall. \textcolor{black}{The Precision is slightly lower than that of the zero-shot method, while the F1-score is comparable to the zero-shot method.}
    \item The increase in ``Insightful'' arguments and the decrease in ``Helpful'' arguments ensure that the DR generated by LLM-based agents provides a comprehensive perspective on the architecture decision while minimizing the inclusion of low-relevance arguments.
    \item The DR generated by LLM-based agents includes the fewest ``Uncertain'' and ``Misleading'' arguments, primarily due to the modifications made by \textit{Aspect\_Reviewer} agent and the analysis provided by \textit{Trade-off\_Analyst} agent.
\end{itemize}
\end{tcolorbox}

\begin{table}[htbp]
\caption{Comparison of evaluation results from the three prompting strategies}
\label{tab:comparison}
\centering
\small
\begin{tabular}{m{3.5cm}ccc}
\hline
\rowcolor{lightgray} & \textbf{zero-shot} & \textbf{CoT} & \textbf{LLM-based Agents} \\ \hline
\textbf{Precision} & \textbf{\underline{0.278}} & 0.237 & 0.267 \\ 
\textbf{Recall} & 0.627 & 0.684 & \textbf{\underline{0.715}}  \\ 
\textbf{F1-score} & 0.381 & 0.351 & \textbf{\underline{0.389}}\\ 
\textbf{Argument Count} & 455 & \textbf{\underline{576.6}} & 531.4\\ \hline
\textbf{I/(I+H+U+M)} & \textbf{\underline{27.44\%}} & 23.68\% & 26.71\% \\ 
\textbf{H/(I+H+U+M)} & 64.45\% & \textbf{\underline{69.42\%}} & 67.58\% \\ 
\textbf{U/(I+H+U+M)} & 4.87\% & 4.57\% & \textbf{\underline{4.12\%}}  \\ 
\textbf{M/(I+H+U+M)} & 3.24\% & 2.33\% & \textbf{\underline{1.59\%}} \\ \hline
\end{tabular}
\end{table}

\subsection{RQ4: How do practitioners perceive LLM-generated DR of architecture decisions in practice?}
\textcolor{black}{As elaborated in Section~\ref{subsec:RQs}, we conducted semi-structured interviews with practitioners to understand their perspectives on LLM-generated DR. To select interview participants, we first invited developers who had engaged in architecture-related discussions in our dataset on SO via email, and also posted the invitation in software development groups on LinkedIn that focus on software architecture. After collecting initial expressions of interest, we screened practitioners based on their professional experience in software development and architecture design, as well as their availability and willingness to attend a semi-structured interview. Based on these criteria, we ultimately selected six developers to participate in the interviews. Each interview took 25 to 30 minutes to complete. Table~\ref{tab:participant_experience} presents the experience of the six practitioners (P1$ \sim $P6) in software development and architecture design. The key results of the interviews are presented below:}

\begin{table}[h]
\caption{\textcolor{black}{Participants' experience in software development and architecture design}}
\centering
\color{black}
\begin{tabular}{ccc}
\hline
\textbf{Participant} & \textbf{Software Development Experience} & \textbf{Architecture Design Experience} \\
\hline
P1 & 3 years & 2 years \\
P2 & 7 years & 2 years \\
P3 & 3 years & 3 years \\
P4 & 25 years & 10 years \\
P5 & 3 years & 2 years \\
P6 & 2 years & 2 years \\
\hline
\end{tabular}
\label{tab:participant_experience}
\end{table}

\begin{itemize}
    \item \textcolor{black}{\textbf{IQ1.1}: Concerning the DR generated by LLMs, two practitioners (P5, P6) chose to ``\textit{generally trust it}''. P2 reported that he could ``\textit{trust it with some caution}'', while P1, P3 and P4 indicated that they could only ``\textit{trust it with substantial human review}''. Overall, all the participants recognized the value of LLM-generated DR, but they also expressed reservations about fully relying on LLMs to recover DR in practice. There are three primary reasons highlighted by the six practitioners for their worries: \textbf{(1) LLMs are unable to fully understand the requirements of the project that architecture decisions must fulfill (P1, P2, P5).} The three practitioners consistently emphasized the complexity of clarifying the project requirements in practice. As a result, even P5 and P6, who demonstrated the highest level of trust in LLM-generated DR, cautioned that ``\textit{LLMs may fail to fully understand specific project requirements}''. \textbf{(2) Human review constitutes an indispensable step in practice to ensure that architectural solutions are well-reasoned and appropriate (P1, P4).} No matter the DR is provided by human experts or generated by LLMs, P4, based on his practical experience, stated that ``\textit{validation should involve discussion among multiple human experts}''. \textbf{(3) Some arguments in the DR generated by LLMs are not accurate or reliable (P3, P4).} From the perspective of P3, the DR generated by LLMs are of ``\textit{mixed quality}''. Therefore, he required more thorough human validation before accepting the generated DR.}
    \item \textcolor{black}{\textbf{IQ1.2}: In our interviews, three types of information were mentioned by the participants as useful for enhancing the trustworthiness of LLM-generated DR: \textbf{(1) Project context (P1$ \sim $P6)}: The context related to the project was most frequently mentioned by the six practitioners. Generally, they believed that providing more detailed project context, such as ``\textit{the goals of the business system, the future development plans, upstream and downstream dependencies, and anticipated workload}'' (P3), could enable LLMs to produce more accurate and contextually appropriate DR. Similarly, according to the experience of P4, prompts containing detailed project-specific context yield high-quality DR. \textbf{(2) The architecture design and DR of other similar projects (P1, P3)}: According to P1, referring to the architecture designs of other similar projects is highly valuable for architects when making architecture decisions, since it is common to ``\textit{adapt successful architectural examples from other projects to meet the specific needs of my own projects}''. \textbf{(3) Domain-specific knowledge (P1, P4)}: In the interviews, P1 and P4 believed that providing domain-specific knowledge could enhance the quality of DR generated by LLMs. However, P3 and P5, although acknowledged the importance of domain-specific knowledge, they argued that such knowledge should not be provided to LLMs by architects. Instead, ``\textit{LLMs should either possess this knowledge inherently or be capable of retrieving it through online access}''.}
    \item \textcolor{black}{\textbf{IQ2.1}: Among the six practitioners, P3 stated that he would \textbf{definitely} use LLMs to generate DR in practice. P1, P5 and P6 indicated that they \textbf{probably} would use LLM-generated DR in practice, while P2 and P4 were \textbf{not sure} about their intentions.}
    \item \textcolor{black}{\textbf{IQ2.2}: In our interviews, practitioners mentioned four scenarios that are suitable for applying LLM-generated DR in practice: \textbf{(1) Comparing different architecture design (P1, P3, P6)}: P1 and P3 considered that the DR generated by LLMs could efficiently help them better identify the key differences and trade-offs between architecture designs by revealing the underlying motivations and objectives behind each design. \textbf{(2) Understanding unfamiliar architecture that exceeds the team's prior experience (P1, P3, P4)}: P1, P3, and P4 concurred that the DR produced by LLMs enables them to rapidly understand architectures that are unfamiliar to them. For instance, P3 noted that the DR produced by LLMs was particularly valuable in aiding comprehension of specific details while studying a new architecture. P4 mentioned that when the company’s business needs to expand into entirely new domains, the DR generated by LLMs can help them quickly identify which architectures are applicable. \textbf{(3) Gaining deeper insight into architecture decisions made in the current system (P2, P5, P6)}: According to P2, once the architectural layers are defined, LLM-generated DR can analyze the design from multiple perspectives based on architectural context. This may lead to more well-founded insights and help developers better understand the details of architecture. P5 and P6 believed that LLM-generated DR is especially valuable in the absence of documentation concerning architecture design and its rationale, as it helps architects gain a more comprehensive understanding of the system architecture. \textbf{(4) Guiding architectural evolution (P4, P5)}: Extending projects with a new functionality module may require modifications or extensions to the existing architecture. P4 noted that in such a scenario, LLM-generated DR can assist in understanding how to update the existing architecture. P5 further emphasized that LLM-generated DR can help architects assess the impact of architectural changes and ensure compatibility between the updated components and the existing architecture.}
    \item \textcolor{black}{\textbf{IQ2.3}: The participants reported the challenges and limitations in four key aspects: \textbf{(1) For architects, collecting and providing sufficient project context for architecture problems to LLMs is often a complex and cost-inefficient process (P1$ \sim $P6).} All practitioners mentioned their concerns about the efficiency of collecting architecture-related context. P2 noted that ``\textit{how to summarize comprehensive project requirements, including scalability, maintainability, and performance, is a challenge task}''. According to P6, when collecting project-specific information becomes overly complex, ``\textit{the cost of using LLMs may outweigh the potential benefits}''. \textbf{(2) Enabling LLMs to understand the complex requirements of specific projects is challenging (P2, P5).} P2 pointed out that ``\textit{whether LLMs can fully understand the complex requirements that architecture decisions must satisfy still needs to be further validated in practice}''. Based on his personal experience, P5 observed that when LLMs are provided with overly long context in the prompt, they may fail to fully comprehend the input, resulting in a decline in the quality of the generated DR.  \textbf{(3) The reliability of LLM-generated DR remains uncertain and cannot be fully trusted by architects (P1, P3).} P1 stated that, compared to LLMs, he would definitely prefer to trust the DR provided by experts who are familiar with the architecture. P3 noted that LLMs may lack sufficient knowledge to effectively analyze the DR of unfamiliar architectures, especially those introduced after the models’ knowledge cut-off date. \textbf{(4) LLMs face inherent restrictions in their output form (P4).} P4 observed that the limited output length of LLMs constrains their ability to generate detailed and comprehensive DR, particularly when the DR involves extensive or complex discussions. In addition, P4 emphasized the critical role of visual representations (e.g., architecture diagrams) in explaining and facilitating the understanding of architecture decisions, which can be potentially facilitated by multimodal LLMs.}
\end{itemize}

\begin{tcolorbox}[colback=gray!20, colframe=gray]
\textbf{\textcolor{black}{Key findings of RQ4}}
\begin{itemize}
    \item \textcolor{black}{Practitioners rated their trust level of LLM-generated DR between ``\textit{generally trust it}'' and ``\textit{trust it with substantial human review}''.}
    \item \textcolor{black}{Providing LLMs with project context, architecture and DR from similar projects, and domain-specific knowledge is considered useful in improving the credibility of generated DR.}
    \item \textcolor{black}{The majority of participants expressed a willingness to use LLMs for generating DR in practice, while two of them remained uncertain.}
    \item \textcolor{black}{``Comparing different architecture design'', ``Learning unfamiliar architecture'', ``Gaining deeper insight into current architecture decision'' and ``Guiding architectural evolution'' are the four main scenarios for applying generated DR in practice, as identified by practitioners.}
    \item \textcolor{black}{``Difficulty in providing context'', ``limited understanding of complex requirements'', ``concerns about reliability'', and ``output constraints'' are the main challenges in applying LLM-generated DR, as reported by practitioners.}
\end{itemize}
\end{tcolorbox}

\section{Discussion} \label{sec:discussion}
%In this section, we discuss the implication based on the results in Section~\ref{sec:results}.

\subsection{Pros \& Cons of Three Prompting Strategies}
\label{subsec:pros&cons}
\noindent \faLightbulbO \quad \textbf{Zero-shot can generate effective DR with single-round dialogue \textcolor{black}{and is commonly employed by practitioners;} however, its key limitation lies in the lack of comprehensive analysis and the inclusion of some irrelevant content.}

Zero-shot is a basic prompting strategy that provides LLMs with the context of architecture decisions and basic instructions to generate DR, requiring only one single LLM invocation. According to the average performance presented in Table~\ref{tab:comparison}, the DR generated with zero-shot method demonstrates the highest Precision (0.278) and an F1-score (0.381) comparable to that of LLM-based agents. For other metrics, the zero-shot approach also shows no significant gap compared to other methods. 
% To some extent, the generated DR align with human expert insights and offer valuable analysis from different perspectives. 
The performance of the zero-shot method indicates that LLMs are capable of generating DR for architecture decisions without additional training. \textcolor{black}{Besides, practitioners (P2, P3, and P4) emphasized that, when provided with sufficient project context, the zero-shot method can capture fine-grained rationale for architectural issues of specific modules or design concerns. Consequently, the zero-shot method is regarded as practical and widely adopted by practitioners for asking DR in specific architectural contexts.}

However, the DR generated through the zero-shot method rarely identifies the potential issues of architecture decisions for analysis. For the five LLMs, only 16\%  of the DR generated by the zero-shot method includes an analysis of architectural weaknesses. Therefore, the arguments provided in the DR generated by the zero-shot method may lack comprehensive analysis. For instance, the DR generated for \href{https://web.archive.org/web/20250622085527/https://stackoverflow.com/questions/77745340/how-to-make-two-projects-with-incompatible-net-versions-talk}{SO \#77745340} by gpt-3.5-turbo incorrectly judges the performance of the architecture decision due to overlooked IPC overhead, as mentioned in Section~\ref{subsubsec:CoT_manual}. The lack of analysis of potential architectural weaknesses may result in skewed conclusions in arguments, which contributes to the highest proportion of ``Misleading'' arguments in the DR generated with the zero-shot method. In addition, the prompt template for the zero-shot method does not specify any requirements for the output. As a result, the analysis of DR generated by the zero-shot method is unpredictable and may occasionally include content unrelated to the DR. For example, the DR generated by gemini-1.0-pro for \href{https://web.archive.org/web/20250623051817/https://github.com/razshare/sveltekit-sse/issues/45}{Issue \#45} also suggests other architecture decisions to mitigate ``thundering-herd scenarios'', where multiple clients or processes simultaneously compete for a shared resource. Such information in the output is unrelated to the DR and may not align with the expectations of software architects regarding the generated content. \textcolor{black}{Therefore, to ensure the DR generated through the zero-shot method aligns with the expected structure, it is recommended to explicitly provide an output template in the prompt during practical use.}

\noindent \faLightbulbO \quad \textbf{CoT enhances the comprehensiveness of DR by incorporating trade-off analysis compared to the zero-shot method, \textcolor{black}{and can potentially accommodate different scenarios;} however, it lacks in-depth analysis for individual arguments.}

The CoT method instructs LLMs to generate DR step by step according to the provided template. All five LLMs correctly follow the instructions in the CoT template during DR generation, outlining the potential advantages and disadvantages of architecture decisions for the trade-off analysis in the final step. Compared to the zero-shot method, the CoT method enhances the reasoning capabilities in DR generation, enabling more thorough analysis during the trade-off process and reducing the proportion of ``Misleading'' arguments \textcolor{black}{(2.33\%)} compared to the zero-shot method \textcolor{black}{(3.24\%)}, as explained in Section~\ref{subsubsec:CoT_manual}. \textcolor{black}{According to the results presented in Table~\ref{tab:comparison}, the CoT method achieves higher Recall (0.684) compared to zero-shot (0.627), while its Precision (0.237) and F1-score (0.351) are slightly lower than those of zero-shot and LLM-based agents. Notably, the DR generated by the CoT method contains the highest proportion of ``Helpful'' arguments (69.42\%) among the three approaches, providing architects with richer perspectives for understanding the architecture.} \textcolor{black}{Moreover, our experiments demonstrate that the reasoning process embedded in the CoT template can effectively guide LLMs in both reasoning about and generating DR. Therefore, practitioners may tailor the CoT reasoning process by modifying prompt templates to better accommodate different practical scenarios. For instance, ``comparing different architecture designs'', in which practitioners consider a valuable scenario, is particularly well-suited to the step-by-step reasoning enabled by the CoT method.}

However, some argument points in the DR generated using the CoT templates lack the analytical depth. According to Table~\ref{tab:comparison}, the number of argument points generated by CoT is 26.7\% higher than that generated by the zero-shot method and 8.5\% higher than that generated by LLM-based agents. Although these arguments provide diverse perspectives on understanding architecture decisions, some of them are merely superficial statements without in-depth analysis and reasoning. For example, in the generated DR for \href{https://web.archive.org/web/20250623053207/https://github.com/grpc/grpc-dotnet/issues/2378}{Issue \#2378} by gpt-3.5-turbo with the CoT method, the discussions on ``increased code complexity'' just briefly mention the difficulty of implementing the architecture without a detailed analysis of the specific challenges or potential implications of this added complexity. Similarly, the generated DR also does not provide specific analysis when considering ``future reusability''. Therefore, software architects could struggle to derive valuable insights from such argument points due to the lack of in-depth analysis.

\noindent \faLightbulbO \quad \textbf{LLM-based agents use multi-agent collaboration for information gathering, reasoning, and verification; however, excessive interactions between agents increase conversation rounds with LLMs, and certain agent functionalities remain to be further optimized.}

\textcolor{black}{LLM-based agents represent the most complex approach among the three prompting strategies. The DR generated by this method achieves the highest Recall (0.715) and F1-score (0.389), while containing the fewest ``Uncertain'' (4.12\%) and ``Misleading'' arguments (1.59\%). Although its Precision is slightly lower than that of the zero-shot method, the higher Recall suggests that LLM-based agents capture a greater number of arguments aligned with those of human experts.} According to Fig.~\ref{fig:llmbasedagents}, the reasoning process of LLM-based agents involves two steps: \textit{Aspect\_Analyst} agent analyzes each aspect related to architecture decisions, while \textit{Trade-off\_Analyst} agent provides the final DR based on the analysis of these aspects. By analyzing each architecture-related aspect individually, \textit{Aspect\_Analyst} can provide a more detailed analysis for every aspect, mitigating the issue in CoT where arguments are outlined without sufficient analysis. \textit{Trade-off\_Analyst} agent is tasked with performing a trade-off analysis based on the evaluation of different aspects of architecture decisions, which have been reviewed by \textit{Aspect\_Reviewer} agent beforehand. Therefore, the final DR generated by \textit{Trade-off\_Analyst} agent also includes an analysis of the advantages and disadvantages of architecture decisions, similar to the CoT method, ensuring the comprehensiveness of the generated DR. By leveraging these two layers of reasoning (i.e., \textit{Aspect\_Reviewer} and \textit{Trade-off\_Analyst} agents), LLM-based agents generate DR that better captures key concerns of human experts in architecture decisions than CoT. According to Table~\ref{tab:comparison}, the DR generated by LLM-based agents exhibits the highest Recall (0.715) among the three prompting strategies, which indicates that it includes most arguments that align with those of human experts. Besides, the review process executed by \textit{Aspect\_Reviewer}, which checks and modifies the analysis results of each aspect by \textit{Aspect\_Analyst} agent, leads to a decreased proportion of ``Misleading'' arguments in the generated DR as explained in Section~\ref{subsubsec:agents_manual}.

However, one of the issues of LLM-based agents in generating DR is the increased conversation rounds with LLMs. The generation of one DR may require up to 20 rounds of conversation with the LLMs, which is \commentword{significantly} greater than required by the zero-shot and CoT methods. In addition, some functions of the agents can be further optimized. For instance, it is challenging to assess the value of external knowledge provided by \textit{Information\_Collector} agent. Although certain DR, such as the one for \href{https://web.archive.org/web/20250623052901/https://github.com/k2-fsa/icefall/issues/1517}{Issue \#1517} generated by mistral-7B, explicitly integrates ``recent findings'' to support the arguments, the impact of external knowledge is difficult to discern in most cases. Therefore, \textit{Aspect\_Analyst} could incorporate functionality for tracking external knowledge in order to clearly identify which arguments are derived from what external knowledge. Furthermore, compared to human experts, the arguments in the LLM-generated DR typically lack concrete code examples. Therefore, equipping \textit{Information\_Collector} with the ability to automatically retrieve open-source project code or code from Q\&A sites relevant to the architecture problem, and enabling \textit{Aspect\_Analyst} to generate context-based code examples, would help LLM-based agents to provide DR closer to practical development scenarios. \textcolor{black}{Moreover, when applying the LLM-based agent system to a specific project, customizing the capability of \textit{Information\_Collector} to extract architecture-related information in depth from that project is of practical significance, as it alleviates the burden on architects by reducing the need to manually gather project context. This challenge was most frequently highlighted by practitioners for applying LLM-generated DR in practical scenarios.}

\subsection{Strengths and Limitations of LLM-Generated DR}
\faLightbulbO \quad \noindent \textbf{\textit{Strength 1:} LLMs are capable of identifying the key factors of DR considered by human experts.}

During the evaluation process, the arguments of DR provided by human experts are regarded as ground truth to quantitatively evaluate the accuracy metrics of LLM-generated DR. Table~\ref{tab:comparison} shows that LLMs exhibit an average Recall higher than 0.6 with all three prompting strategies, in which LLM-based agents even exceed 0.7. This result reveals that most of the key points considered by human experts when making architecture decisions are identified in LLM-generated DR. To enable LLMs to perform multidimensional analysis from different aspects for each architecture problem, we did not explicitly specify in the prompt which perspectives should be focused on. Therefore, the Recall results suggest that LLMs have the potential to reason the key factors of DR based on the specific context of architecture problems and their corresponding architecture decisions.

\noindent \faLightbulbO \quad \textbf{\textit{Strength 2:} LLMs tend to provide more arguments from diverse perspectives in DR than human experts.}

\textcolor{black}{The number of arguments in the LLM-generated DR is \commentword{significantly} higher than that provided by human experts. For each architecture decision, human experts provide an average of 1.98 arguments in DR, while LLMs generate an average of 4.55, 5.77, and 5.31 arguments using the three prompting strategies (i.e., zero-shot, CoT, LLM-based agents), respectively. Despite covering most of the arguments provided by human experts, the greater number of arguments in LLM-generated DR leads to a Precision below 0.3 and an F1-score below 0.4.} \textcolor{black}{Nevertheless, this result does not suggest that the quality of the DR generated by LLMs is low. Accuracy metrics are calculated based on the human experts' rationale as ground truth; however, these expert-provided DR are not necessarily the only correct answers and may be incomplete. In our IHUM-category evaluation, the majority of arguments that were not aligned with those provided by human experts were still considered ``Helpful''.} \textcolor{black}{The presence of more than 60\% ``Helpful'' arguments in LLM-generated DR} suggests that LLMs typically analyze a broader range of aspects of DR for architecture decisions compared to human experts. Human experts may prioritize the feasibility of architecture implementation and its alignment with the project requirements, potentially overlooking other critical considerations when analyzing DR. For example, in \href{https://web.archive.org/web/20250622084557/https://stackoverflow.com/questions/77579165/this-is-about-the-repository-layer-when-implementing-a-clean-architecture-using}{SO \#77579165}, the expert suggested ``\textit{creating a concrete repository class and mocking it for testing in Riverpod and Flutter}'', simplifying implementation by synchronizing two files for efficiency and clear responsibility separation. In contrast, the DR generated by llama3-8B expands on this by further emphasizing the separation of concerns, testability, and reusability. \textcolor{black}{Thus, an additional value of LLM-generated ``Helpful'' arguments is to complement the DR provided by human experts and help architects to understand architecture decisions from diverse perspectives. Such advantage was also mentioned by some practitioners during our interviews. For instance, P2 noted that ``\textit{Compared to practitioners, LLMs are typically able to conduct more comparative analyses from multiple perspectives, thereby providing more reasonable DR}''.} \textcolor{black}{Besides, some ``Helpful'' arguments generated by LLMs highlight potential disadvantages of architecture decisions, which reveal trade-offs that were overlooked by human experts. For example, in \href{https://web.archive.org/web/20250623050023/https://github.com/pointfreeco/swift-composable-architecture/discussions/2792}{Discussion \#2792}, the human expert focused primarily on the benefits of modular design, including improved decoupling, enhanced build efficiency, and the prevention of circular dependencies. In contrast, the DR generated by gpt-3.5-turbo using the LLM-based agents strategy offers a balanced analysis of the pros and cons for each key argument, highlighting both the architectural advantages and the potential overhead associated with management and coordination. When explicitly guided to perform trade-off reasoning through CoT and LLM-based agents strategies, LLM-generated DR can typically reveal architectural trade-offs through those ``Helpful'' arguments that human experts overlooked.}

\noindent \faLightbulbO \quad \textbf{\textit{Limitation 1:} DR generated by LLMs may lack the practical project experience and context-specific details that human experts provide.}

While LLMs can generate DR from perspectives aligned with those of human experts, they often lack the depth of practical experience and contextual understanding, which are essential characteristics of human expertise. For example, in \href{https://web.archive.org/web/20250623052343/https://github.com/valkey-io/valkey/issues/65}{Issue \#65}, ``\textit{Debian}'' was chosen as the operating system for the container image as an architecture decision. This decision was explained by human experts who not only emphasized Debian's advantages, but also considered the time loss associated with code refactoring incurred due to migrations from other Linux distributions (e.g., \textit{Amazon Linux 2}, \textit{Oracle Linux}, and \textit{Universal Base Image}) to Debian-based systems. This practice-based analysis is more persuasive for supporting the architecture decision, whereas the DR generated by LLMs typically lacks such experience-driven arguments. In \href{https://web.archive.org/web/20250623042641/https://github.com/developmentseed/lonboard/discussions/442}{Discussions \#442}, while the DR generated by LLMs also addressed the complexity of implementing the architecture at the code level, human experts were able to explicitly identify specific potential technical barriers according to the architecture context. \textcolor{black}{Therefore, besides the capability of LLMs in generating DR, the quality of DR is also influenced by the contextual information provided. However, in this study, SO posts usually offer only the necessary context needed to describe a specific architecture problem, without giving a full project background. Although GitHub issues and discussions are associated with specific projects and allow additional background information to be retrieved from the corresponding repositories, the key challenge lies in automatically determining which parts of the context are relevant to a given architecture problem, since manual filtering is highly time-consuming and labor-intensive. Additionally, regarding domain-specific knowledge, the large amount of potentially relevant background information makes it difficult to decide what is most useful for LLMs in generating DR. As a result, determining how to effectively use relevant project context and domain-specific knowledge is crucial for enabling LLMs to exhibit context-aware insights like human experts when analyzing arguments for DR, thereby improving the quality of the generated DR.}

\noindent \faLightbulbO \quad \textbf{\textit{Limitation 2:}  The proportion of ``Uncertain'' arguments in the DR generated by LLMs is difficult to reduce, primarily due to constraints in the provided context and insufficient domain-specific details.}

The proportion of ``Uncertain'' arguments is 4.87\%, 4.57\%, and 4.12\% with the three prompting strategies, respectively. Compared to the zero-shot method, neither the CoT method nor LLM-based agents \commentword{significantly} reduce the proportion of ``Uncertain'' arguments. A potential reason is the insufficient contextual information related to the architecture. Although we provided as much detailed contextual information as possible collected from the data extraction phase, the available information for LLMs to generate DR remains limited. For instance, regarding architecture problems in GitHub projects, LLMs lack critical insights that project developers possess, such as the previous versions and specific code contexts. Moreover, the challenges that architecture design faces in real-world scenarios remain unknown to LLMs. As a result, the ``Uncertain'' arguments proposed by LLMs require validation in practical production environments. However, ``Uncertain'' arguments may also offer valuable insights, encouraging architects to further analyze and consider them in their project context.
% \noindent \textbf{Instead of having LLMs directly generate the DR, explicitly prompting LLMs to perform trade-off analysis can significantly enhance their reasoning capacity.}

\subsection{Implications from Interviews}
\noindent \faLightbulbO \quad \textbf{\textcolor{black}{\textit{Implication 1:} Practitioners consider LLMs as an efficient assistant in generating and recovering DR for architecture decisions, rather than a fully trustworthy expert.}}

\textcolor{black}{By inquiring into the trustworthiness and applicability of LLM-generated DR, we found that practitioners regarded DR as a useful reference and were able to identify appropriate scenarios for practical application. However, practitioners typically regarded LLM-generated DR as a supplementary aid to support their architectural understanding, rather than as authoritative expert advice. Specifically, half of the practitioners in our interviews stated that trusting LLM-generated DR requires substantial human review. Other practitioners expressed a higher trust level, but none would ``\textit{fully trust it without human validation}''. From the perspective of practitioners, LLM-generated DR is seen as less trustworthy than that provided by experienced human experts. For this reason, P5 noted that he was ``\textit{not sure}'' whether he would rely on LLMs' DR when working within a domain his team is already familiar with. To enhance practitioners’ trust in LLM-generated DR, two key challenges need to be addressed. First, it is essential to demonstrate that LLMs are capable of understanding complex project-related contexts - one of the primary concerns raised by practitioners during the interviews. Without such capability, DR generated by LLMs may be perceived as simplistic, thereby limiting their value in supporting practitioners' in-depth understanding of architecture design. Second, building trust in LLM-generated DR requires extensive empirical validation in practical software development settings. Practitioners emphasized the importance of evaluating how LLM-generated DR performs in real development processes, particularly in terms of accurately revealing the rationale of architecture decisions and supporting meaningful reasoning across diverse project contexts.}

\noindent \faLightbulbO \quad \textbf{\textcolor{black}
{\textit{Implication 2:} Providing LLMs with project-specific context is highly beneficial, yet challenging to achieve due to the associated costs.}}

\textcolor{black}{During our interviews, all participants consistently emphasized that providing LLMs with detailed project context significantly enhances the quality and usefulness of the generated DR. However, they also pointed out that the cost of collecting and providing such contextual information is non-negligible for architects, which is also the most frequently mentioned challenge across the interviews. For example, P5 noted that ``\textit{if the process of collecting project context is too burdensome, it could outweigh the benefits of using LLMs to generate DR}''. To address this issue, a potential solution is to further enhance the capability of \textit{Information\_Collector} agent to extract project-specific context. In this study, \textit{Information\_Collector} agent was designed to dynamically retrieve contextual information using a search engine, in order to support diverse architectural scenarios for different projects. However, in real-world software development, practitioners typically focus on understanding the architecture of a single project. Therefore, it is feasible to customize \textit{Information\_Collector} agent for a specific project by targeting the artifacts of the project (e.g., code, documentation, and configuration files). By aligning the extraction process with the architect's specific information needs, \textit{Information\_Collector} agent can automatically retrieve key architecture-relevant context. This capability is crucial for improving both the quality and practical utility of LLM-generated DR.}

\noindent \faLightbulbO \quad \textbf{\textcolor{black}{\textit{Implication 3:} Enhancing LLM-based agents with visual output capabilities can facilitate comprehension of DR.}}

\textcolor{black}{In our interviews, P1 and P4 commented on the textual form of DR generated by LLMs. P1 noted that, compared to DR provided by human experts, the longer textual outputs generated by LLMs imposed a certain cognitive load, which made him ``\textit{somewhat reluctant to read them in full}''. In addition, P5 argued that the visual form may be more effective than textual explanations in some cases, which enables architects to ``\textit{quickly comprehend the architecture and the rationale behind it}''. Similarly, according to the experiment conducted by Shahin \textit{et al.} \cite{shashin2014do}, visualization of architectural design decisions and their DR can improve the understanding of architecture design. Consequently, a promising direction for our future work lies in equipping LLM-based agents with the ability to generate visual representations of architecture decisions with their corresponding DR, which can better assist architects in comprehending DR and mitigating cognitive burden.}

\section{Threats to Validity} \label{sec:validity}
In this section, we discussed the threats to validity following the guidelines by Runeson and H{\"o}st \cite{runeson2009guidelines}. It should be mentioned that internal validity is excluded, as we did not explore the causal relationships between variables and outcomes.

\textbf{Construct Validity}.
The main threat to the construct validity of this study is the evaluation methods we employed to assess the quality of LLM-generated DR. Although we selected architecture decisions with DR provided by human experts in our dataset, it is important to note that there is typically no universally definitive DR for a given architecture decision. Therefore, we employed two complementary methods for evaluating the generated DR: accuracy metrics and the IHUM-category. For accuracy metrics, we compared the arguments in the generated DR with those provided by human experts to calculate the Precision, Recall, and F1-score value. Since the DR provided by human experts is used to explain architecture decisions, the values of the accuracy metrics can reflect the capability of LLMs to recover the DR for architecture decisions. However, accuracy metrics considered the DR provided by human experts as the ground truth, overlooking the arguments in LLM-generated DR that may differ from those of human experts. Therefore, we proposed the IHUM-category to qualitatively assess the arguments in the generated DR, categorizing them into those that align with the human experts' arguments (``Insightful''), those that differ but offer valid perspectives (``Helpful''), those that are ambiguous (``Uncertain''), and those contain erroneous information (``Misleading''). We evaluated the quality of LLM-generated DR by using both methods simultaneously, aiming to assess generated DR as comprehensively as possible.
\textcolor{black}{Another potential threat to construct validity arises from \textit{Information\_Collector} agent, which gathers relevant knowledge via search engines. Although we excluded URLs that overlapped with those in our dataset to avoid direct access to original data entries, content mirroring on other websites (e.g., from SO) may still result in unintended information leakage. To fully avoid this issue, search engine date-range restrictions can be used to ensure all retrieved information predates the creation of the architecture problem. However, this approach may also limit the external knowledge \textit{Information\_Collector} agent can obtain from search engines, as some potentially useful recent information might be filtered out. In our future work, we can optimize the filtering strategy to effectively minimize information leakage while maximizing the capture of potentially valuable external knowledge for \textit{Information\_Collector} agent.}

\textbf{External Validity}.
The primary threat to the external validity is the selection of data sources. In our dataset, we chose the posts from SO, as well as the issues and discussions from GitHub to maximize the diversity of data sources. SO is a popular Q\&A community where developers engage in discussions, including a wide range of architecture-related issues~\cite{musengamana2023characterizing}. GitHub Issues are used for tracking bugs, feature requests, and raising potential issues, while GitHub Discussions facilitate project-related discussions and knowledge sharing. Both the two sources on GitHub can provide architecture decisions from practical projects. Therefore, the three data sources ensure, to a certain extent, the diversity of our dataset. However, we acknowledge that our dataset is not comprehensive and does not cover all architecture problems. %Besides, we only selected the data created after the cut-off date of the five LLMs to prevent the knowledge leakage issue, which also resulted in overlooking some valuable data. 
Another threat to the external validity is the selection of LLMs. We chose five widely used open-source and closed-source LLMs that have demonstrated strong performance in other SE research. However, the performance of these five LLMs may not fully represent the state-of-the-art capabilities of LLMs in generating DR, as other models with potentially superior performance may not be included.  \textcolor{black}{Moreover, based on the results of five selected LLMs, it remains difficult to reveal the relationship between the quality of generated DR and model-specific attributes, such as parameter size, model architecture, or release date. On the one hand, the outputs of LLMs are inherently difficult to interpret; on the other hand, the number of models included in our experiments is limited, making it difficult to perform a generalizable analysis. For example, gpt-4-0613 achieves the best performance in terms of Recall, as well as in generating ``Helpful'' arguments while avoiding ``Uncertain'' and ``Misleading'' ones. However, due to the lack of access to gpt-4-0613's internal mechanisms and training data, as well as the absence of comparative experiments (e.g., control for model size), it is challenging to rigorously explain the reasons behind its superior performance.} \textcolor{black}{In addition, as discussed in Section~\ref{subsec:LLMselection}, there is a trade-off between ensuring a sufficiently large dataset and selecting the most recent LLMs. As a result, the five selected LLMs in our study were released over a span of one year and one month. Although our experimental results did not show an advantage of the more recently released model (i.e., llama3-8B) over others, the inconsistency in the release periods of the LLMs remains a potential threat to external validity.} \textcolor{black}{Therefore, we plan to continue exploring the capabilities of a broader range of models released across different time periods for generating DR in our future work, and to design controlled comparative experiments to systematically investigate how model-specific attributes - such as size, architecture, and training data - influence DR quality.} 

\textbf{Reliability}. 
%refers to the degree to which a research method can consistently produce reliable and reproducible results. 
Since the dataset construction and results evaluation process were conducted manually, there are potential to introduce personal bias. To reduce this threat, the first and fourth authors conducted pilot experiments before data labeling, extraction, and IHUM-category classification. During the data labeling and evaluation of the results (i.e., using accuracy metrics and IHUM-category), the consistency between the first and fourth authors was evaluated using the Cohen's Kappa coefficient, yielding values of 0.834, 0.863, and 0.738, respectively. The Cohen's Kappa value indicates an almost perfect agreement between the two authors on data labeling and accuracy metrics, and substantial agreement on the IHUM-category classification results \cite{landis1977theMO}. If any disagreement arose during the pilot and formal process, the second and third authors were involved in reaching a consensus. The constructed dataset and evaluation results were reviewed multiple times by the four authors (the first to fourth authors) to ensure that they aligned with the pre-established criteria. \textcolor{black}{However, since both data labeling and evaluation were conducted by the authors, the Kappa values - although indicative of almost perfect agreement between the first and fourth authors - may not fully guarantee objectivity. While the Kappa statistic measures the degree of agreement among annotators, it does not account for potential systematic biases. For instance, if all annotators exhibit similar interpretive tendencies, the high agreement may reflect shared subjective perspectives rather than genuine objectivity. To reduce such a threat to reliability and foster reproducibility, we have provided the pilot and formal results of data labeling and evaluation in the dataset of this study~\citep{dataset}, including the pilot and formal results of data labeling and evaluation, enabling other researchers to replicate and validate our findings.}

%\textbf{Internal Validity}. 
%The temperature of LLMs is a parameter that determines whether the output is more random and creative or more predictable. In this study, we set the temperature to 1.0, which is typically a default setting, aiming for a balance between randomness and determinism. However, varying the parameter settings can potentially affect the outputs of LLMs, leading to the generation of different DR. Therefore, the influence of temperature on the quality of LLM-generated DR requires further investigation. Moreover, the size of LLMs is also a threat to internal validity. The three selected closed-source models (i.e., gpt-3.5-turbo, gpt-4-0613, and gemini-1.0-pro) do not disclose their exact parameter counts. For the two selected open-source models (i.e., Llama3-8B and Mistral-7B), we chose versions with similar parameter sizes. Conducting experiments on all versions with different sizes (e.g., Llama3-70B and Mistral-small) was not feasible due to the high computational and time costs. Although the performance of the five LLMs in generating DR did not show significant differences under different prompting strategies, the influence of parameters on their output requires further investigation.

\section{Related Work}\label{sec:relatedwork}

\subsection{Design Rationale in Software Architecture}
Several studies focused on DR in software architecture. Van der Ven \textit{et al.} \cite{vanderVen2006design} indicated that explicitly modeling design decisions in software architecture can bridge the gap between rationale management and architecture artifacts, as it enables a close integration of rationale management with the architecture. Tang \textit{et al.} \cite{tang2006asurvey} conducted a survey targeting practitioners to explore their perceptions of the value of DR and to understand how they utilize and document the background knowledge associated with their architecture decisions. In another study, Tang \textit{et al.} \cite{tang2007rationale} introduced a rationale-based architecture model that integrates design rationale, design objects, and their relationships to address the issue of design rationale often being undocumented or unstructured. Davide \textit{et al.} \cite{davide2008value} proposed a new approach to Design Decision Rationale Documentation (DDRD), which tailors the documentation based on its intended use or purpose. In another study, Davide \textit{et al.} \cite{davide2013value} conducted an empirical study through two controlled experiments and proved that the value of a Design Rationale Documentation (DRD) information item depends on its category (e.g., Assumptions, Related Requirements) and the activity it supports. Soliman \textit{et al.} \cite{soliman2024exploring} analyzed 156 architectural emails to identify the design rationale used in mailing lists and their relationship with decision types. They identified nine types of design rationale, six relationships between rationale types, and three relationships between decision and rationale types. Zhao \textit{et al.} \cite{zhao2024drminer} developed DRMiner, to automatically mine latent design rationale from the discussion of developers in open-source communities. Meanwhile, design rationale is a core element of architectural design decisions~\cite{shahin2009architectural}. Dhar \textit{et al.} \cite{dhar2024can} explored the feasibility of using LLMs to generate architectural design decisions based on contextual information, providing preliminary evidence of LLMs' ability to assist in generating design rationale. Building on this line of work, Dhar \textit{et al.} \cite{dhar2025drafting} further investigated few-shot prompting, retrieval-augmented generation, and fine-tuning to enhance LLM-based generation of architecture decisions, and proposed DRAFT, which outperforms prior approaches in generating architecture decisions on a dataset of 4,911 Architecture Decision Records (ADRs). Díaz-Pace \textit{et al.} \cite{díaz-pace2024helping} proposed ArchMind, a generative AI-based design assistant that leverages architectural knowledge and system information to help novice architects select and assess alternative architecture decisions, generating well-justified ADRs in a classroom experiment.

\subsection{LLM-based Understanding of Software Artifacts}
Numerous studies have been conducted to explore the capability of LLMs to understand various software artifacts. Xie \textit{et al.} \cite{xie2024leveraging} evaluated the ability of LLMs to generate software specifications from software comments or documentation using few-shot learning. They compared the performance of 13 state-of-the-art LLMs with traditional methods across three publicly available datasets. Additionally, they conducted a comparative analysis of the failure cases from LLMs and traditional approaches, highlighting their respective strengths and weaknesses. Nam \textit{et al.} \cite{nam2024using} developed an LLM-based conversational user interface to assist users in understanding code. The results show that using their plugin is more effective in code completion than relying on web searches. Shaike \textit{et al.} \cite{shaik2024s3s3llm} presented S3LLM, a framework based on LLMs, to understand large-scale scientific software using source code, code metadata, and summarized information from textual technical reports. 
% Yin \cite{yin2024prosconsevaluatingchatgpt} evaluated ChatGPT's ability to detect vulnerabilities in code and found that existing state-of-the-art methods generally outperform ChatGPT in software vulnerability detection. 
Pex \cite{chaudron2024exploring} studied the application of LLMs in software explanation and proposed the FLASE tool, which combines RAG and knowledge graphs, to enhance developers' understanding of software systems. Pan \textit{et al.} \cite{pan2025codellmsunderstanddesign} conducted an empirical study to evaluate the capability of LLMs to understand design patterns. Franciscatto Guerra \textit{et al.} \cite{guerra2025acccessing} investigated the ability of an LLM to comprehend, replicate, and create structures within the intricate VIPER architecture, a design pattern used in iOS application development. Their results emphasize the potential of LLMs to lower development costs, as well as the challenges the LLMs face in being effectively applied to real-world software design situations. Casillo \textit{et al.} \cite{casillo2025towards} constructed a dataset of 45,945 commits, each accompanied by a rationale explaining the code changes. They then trained a model for generating a rationale for code changes, highlighting the challenges in automating this process. Soliman \textit{et al.} \cite{soliman2025do} evaluated LLMs' understanding of software architecture by comparing their responses to a predefined ground truth. They found that while GPT offers initial insights, expert validation is still needed for reliable results of architecture knowledge. In their study, LLMs were also employed to generate DR for architecture decisions in HDFS. 

\subsection{Conclusive Summary}
Previous studies \cite{vanderVen2006design, tang2006asurvey, tang2007rationale} have highlighted the significant value of DR in software architecture. Besides, the application of LLM-based methods for understanding software artifacts \cite{xie2024leveraging, chaudron2024exploring, soliman2025do} has demonstrated the potential of LLMs in comprehending software architecture and generating DR. However, there has been limited focus on examining the capability of LLMs to generate DR for architecture decisions. Although the study of Soliman \textit{et al.} \cite{soliman2025do} covered querying LLMs about DR related to the software architecture of HDFS, they mainly focused on examining whether LLMs possess specific architectural knowledge using the zero-shot method, rather than on the ability of LLMs to recover DR according to architecture problems and decisions. Dhar \textit{et al.} \cite{dhar2024can, dhar2025drafting} concentrated on improving the quality of architecture decisions generated by LLMs through various prompting strategies; however, their study did not provide a systematic examination of the quality of the corresponding DR. Díaz-Pace \textit{et al.} \cite{díaz-pace2024helping} proposed a generative AI–based design assistant (ArchMind) aimed at supporting novice architects. While their results suggest that LLMs can produce architecture decisions with reasonable rationale, the study did not extend to a systematic analysis of the generated DR or to an explicit assessment of LLMs' ability to recover or generate DR. %Therefore, they did not provide any context related to the HDFS architecture in prompts and only used the zero-shot method for prompting.
In our research, we placed greater emphasis on examining the capability of LLMs to generate DR for diverse real-world architecture problems and corresponding architecture decisions by collecting data from three sources (i.e., SO, GitHub Discussion, and GitHub Issues). Furthermore, we employed two complementary methods (accuracy metrics and the IHUM-category) to evaluate how the optimization of prompting strategies can enhance the quality of the generated DR by employing three different prompting strategies (i.e., zero-shot, CoT, and LLM-based agents).

\section{Conclusions and Future Work} \label{sec:conclusion}
In this study, we analyzed the performance of LLMs in generating and recovering DR for architecture decisions. Our dataset consists of 50 SO posts, 25 GitHub issues, and 25 GitHub discussions that contain architecture decisions and corresponding DR. We used three prompting strategies (i.e., zero-shot, CoT, and LLM-based agent system) to instruct five selected LLMs to generate DR according to provided architecture problems and decisions. Finally, we evaluated the LLM-generated DR by both quantitative accuracy metrics (Precision, Recall, and F1-score) and qualitative analysis (IHUM-category). Our results show that, with the DR provided by human experts as the ground truth, the Precision of LLM-generated DR across the three prompting strategies ranges from 0.267 to 0.278, the Recall spans from 0.627 to 0.715, and the F1-score varies between 0.351 and 0.389. In addition, 64.45\% to 69.42\% of the arguments of generated DR not mentioned by human experts are also helpful, 4.12\% to 4.87\% of the arguments have uncertain correctness, and 1.59\% to 3.24\% of the arguments are potentially misleading. Compared to the zero-shot method, the CoT method generates DR with more arguments and explicitly considers trade-offs by weighing both advantages and disadvantages. However, the DR generated by the CoT method may lack in-depth analysis for individual arguments. The LLM-based agent system for DR generation acquires capabilities in information gathering, two layers of reasoning (i.e., \textit{Aspect\_Reviewer} and \textit{Trade-off\_Analyst} agents), and verification through multi-agent collaboration. Compared to the zero-shot and CoT methods, the DR generated by LLM-based agents is more aligned with the perspectives of human experts, and also contains fewer ``Misleading'' arguments. Additionally, the five LLMs can identify most of the key factors in the DR considered by human experts with the three prompting strategies and provide more comprehensive arguments. However, the five LLMs often struggle to provide analyses grounded in practical project experience and context-specific details as human experts do, and tend to include some ``Uncertain'' arguments, which are challenging to mitigate with the three employed prompting strategies.

\textcolor{black}{Based on our semi-structured interviews, practitioners found LLM-generated DR valuable, but some noted that it might require substantial human review to be fully trusted. They emphasized that providing project context, architectures and DR from similar projects, and domain-specific knowledge could enhance the credibility of the generated DR. Most practitioners expressed willingness to adopt LLMs for DR generation in practice, and the participants identified four typical application scenarios for applying generated DR: comparing alternative architecture designs, learning unfamiliar architectures, gaining deeper insights into current architecture decisions, and guiding architectural evolution. At the same time, they highlighted four major challenges, which include the difficulty of providing LLMs with sufficient project context, limited understanding of complex requirements by LLMs, concerns about the reliability of LLM-generated DR, and the restriction of LLM's outputs in textual form.}

With the evolution of LLMs, we plan to continuously explore the performance of emerging LLMs in DR generation. Additionally, according to the limitations observed in the five selected LLMs for DR generation, we aim to further optimize our LLM-based agent system. \textcolor{black}{To be specific, we intend to enhance \textit{Information\_Collector} agent to automatically gather project-relevant context, including code, documentation, and configuration files, in order to reduce the burden on practitioners while providing \textit{Aspect\_Analyst} agent with the necessary project context for generating DR.} We also aim to augment \textit{Aspect\_Analyst} agent with the ability to identify the external knowledge underlying the generated arguments, as well as automatically generate code examples based on the arguments, to help LLM-based agents provide DR closer to practical development scenarios. \textcolor{black}{Moreover, our future work will explore equipping LLM-based agents with the ability to generate visual representations of DR, which may facilitate comprehension and alleviate cognitive load for practitioners.}

\section*{Data Availability}
The replication package for this work has been made available at \cite{dataset}.

\begin{acks}
This work has been partially supported by the National Natural Science Foundation of China (NSFC) with Grant No. 92582203, 62172311, 62402348, and the Major Science and Technology Project of Hubei Province with Grant No. 2024BAA008. The numerical calculations in this paper have been done on the supercomputing system in the Supercomputing Center of Wuhan University.
\end{acks}

\bibliographystyle{ACM-Reference-Format}
\bibliography{basebib}

@String{Computing = "Computing" }

@String{Springer = "Springer-Verlag" }

@article{delile2023evaluating,
    author={Zack Delile and Sean Radel and Joe Godinez and Garrett Engstrom and Theo Brucker and Kenzie Young and Sepideh Ghanavati},
    title={Evaluating Privacy Questions From Stack Overflow: Can ChatGPT Compete?},
    journal={arXiv preprint arXiv:2306.11174},
    year={2023}
}

@inproceedings{dhar2024can,
  title={Can LLMs Generate Architectural Design Decisions: An Exploratory Empirical Study},
  author={Dhar, Rudra and Vaidhyanathan, Karthik and Varma, Vasudeva},
  booktitle={Proceedings of the 21st IEEE International Conference on Software Architecture (ICSA)},
  year={2024},
  pages={79--89},
  organization={IEEE}
}

@article{musengamana2023characterizing,
  title = {Characterizing architecture related posts and their usefulness in Stack Overflow},
  journal = {Journal of Systems and Software},
  year = {2023},
  author = {Musengamana Jean de Dieu and Peng Liang and Mojtaba Shahin and Arif Ali Khan},
  volume = {198},
  pages = {111608}
}

@article{kabir2024stack,
    author={Samia Kabir and David N. Udo-Imeh and Bonan Kou and Tianyi Zhang},
    title={Is Stack Overflow Obsolete? An Empirical Study of the Characteristics of ChatGPT Answers to Stack Overflow Questions},
    journal={arXiv preprint arXiv:2308.02312},
    year={2024}
}

@manual{babyagi,
    author={Yohei},
    title={BabyAGI},
    year = {2023},
    note={\url{https://github.com/yoheinakajima/babyagi}}
}

@article{gao2024retrieval,
    author={Yunfan Gao and Yun Xiong and Xinyu Gao and Kangxiang Jia and Jinliu Pan and Yuxi Bi and Yi Dai and Jiawei Sun and Meng Wang and Haofen Wang},
    title={Retrieval-Augmented Generation for Large Language Models: A Survey},
    journal={arXiv preprint arXiv:2312.10997v5},
    year={2024}
}

@article{jin2024mare,
    author={Dongming Jin and Zhi Jin and Xiaohong Chen and Chunhui Wang},
    title={MARE: Multi-Agents Collaboration Framework for Requirements Engineering},
    journal={arXiv preprint arXiv:2405.03256},
    year={2024}
}

@inproceedings{jason2022chain,
 author = {Wei, Jason and Wang, Xuezhi and Schuurmans, Dale and Bosma, Maarten and ichter, brian and Xia, Fei and Chi, Ed and Le, Quoc V and Zhou, Denny},
 booktitle = {Proceedings of the 36th Annual Conference on Neural Information Processing Systems (NeurIPS)},
 pages = {24824--24837},
 title = {Chain-of-Thought Prompting Elicits Reasoning in Large Language Models},
 year = {2022}
}

@article{kabir2023answers,
  title={Who Answers It Be er? An In-Depth Analysis of ChatGPT and Stack Overflow Answers to So ware Engineering estions},
  author={Kabir, Samia and Udo-Imeh, David N and Kou, Bonan and Zhang, Tianyi},
  journal={arXiv preprint arXiv:2308.02312},
  year={2023}
}

@article{Sungmin2024evaluating,
  author={Kang, Sungmin and Yoon, Juyeon and Askarbekkyzy, Nargiz and Yoo, Shin},
  journal={IEEE Transactions on Software Engineering}, 
  title={Evaluating Diverse Large Language Models for Automatic and General Bug Reproduction}, 
  year={2024},
  volume={50},
  number={10},
  pages={2677--2694},
}

@article{de2023lessons,
  title={Lessons from Building CodeBuddy: A Contextualized AI Coding Assistant},
  author={de Souza, Cleidson and Neto, Jo{\~a}o Batista and de Souza, Alberto and Gotto, Tarc{\'\i}sio and Monteiro, Edward and others},
  journal={arXiv preprint arXiv:2311.18450},
  year={2023}
}

@misc{gpt-4o,
    author={OpenAI},
    title={Hello GPT-4o},
    note={\url{https://openai.com/index/hello-gpt-4o/}},
    year={2024}
}

@manual{dataset,
  author    = {Xiyu Zhou and Ruiyin Li and Peng Liang and Beiqi Zhang and Mojtaba Shahin and Zengyang Li and Chen Yang},
  title     = {Replication Package of the Paper ``Using LLMs in Generating Design Rationale for Software Architecture Decisions''},
  year      = 2025,
  note      = {\url{https://github.com/Eric0052/LLM4DR}}
}

@misc{stackexchange,
    author={Stack Explorer},
    title={Stack Exchange Data Explorer},
    note={\url{https://data.stackexchange.com/}},
    year={n.d.}
}

@article{barua2014developers,
  title={What are developers talking about? an analysis of topics and trends in stack overflow},
  author={Barua, Anton and Thomas, Stephen W and Hassan, Ahmed E},
  journal={Empirical Software Engineering},
  year={2014},
  volume={19},
  number={3},
  pages={619--654}
}

@article{hong2023metagpt,
  title={MetaGPT: Meta programming for multi-agent collaborative framework},
  author={Hong, Sirui and Zheng, Xiawu and Chen, Jonathan and Cheng, Yuheng and Wang, Jinlin and Zhang, Ceyao and Wang, Zili and Yau, Steven Ka Shing and Lin, Zijuan and Zhou, Liyang and others},
  journal={arXiv preprint arXiv:2308.00352},
  year={2023}
}

@article{liu2024agent,
  title={Agent Design Pattern Catalogue: A Collection of Architectural Patterns for Foundation Model based Agents},
  author={Liu, Yue and Lo, Sin Kit and Lu, Qinghua and Zhu, Liming and Zhao, Dehai and Xu, Xiwei and Harrer, Stefan and Whittle, Jon},
  journal={arXiv preprint arXiv:2405.10467},
  year={2024}
}

@article{widyasari2024codeagent,
  title={CodeAgent: Autonomous Communicative Agents for Code Review},
  author={Ratnadira, Widyasari and Ting, Zhang and Abir, Bouraffa and Walid, Maalej and David, Lo},
  journal={arXiv preprint arXiv:2311.09020},
  year={2024}
}

@inproceedings{hannak2013measuring,
  title={Measuring personalization of web search},
  author={Hannak, Aniko and Sapiezynski, Piotr and Molavi Kakhki, Arash and Krishnamurthy, Balachander and Lazer, David and Mislove, Alan and Wilson, Christo},
  booktitle={Proceedings of the 22nd International Conference on World Wide Web (WWW)},
  pages={527--538},
  year={2013}
}

@article{douze2024faiss,
  title={The faiss library},
  author={Douze, Matthijs and Guzhva, Alexandr and Deng, Chengqi and Johnson, Jeff and Szilvasy, Gergely and Mazar{\'e}, Pierre-Emmanuel and Lomeli, Maria and Hosseini, Lucas and J{\'e}gou, Herv{\'e}},
  journal={arXiv preprint arXiv:2401.08281},
  year={2024}
}

@article{zhang2024using,
  title={Using Large Language Models for Commit Message Generation: A Preliminary Study},
  author={Zhang, Linghao and Zhao, Jingshu and Wang, Chong and Liang, Peng},
  journal={arXiv preprint arXiv:2401.05926},
  year={2024}
}

@article{widjojo2023addressing,
  title={Addressing compiler errors: Stack overflow or large language models?},
  author={Widjojo, Patricia and Treude, Christoph},
  journal={arXiv preprint arXiv:2307.10793},
  year={2023}
}

@article{jacob1960coefficient,
    author={Jacob Cohen},
    title={A Coefficient of Agreement for Nominal Scales},
    journal={Educational and Psychological Measurement},
    year={1960},
    volume={20},
    number={1},
    pages={37--46},
    publisher={Sage}
}

@inproceedings{zhao2024drminer,
    author = {Zhao, Jiuang and Yang, Zitian and Zhang, Li and Lian, Xiaoli and Yang, Donghao and Tan, Xin},
    title = {DRMiner: Extracting Latent Design Rationale from Jira Issue Logs},
    year = {2024},
    booktitle = {Proceedings of the 39th IEEE/ACM International Conference on Automated Software Engineering (ASE)},
    pages = {468--480},
}

@book{bass2021software,
   author = {Bass, Len and Clements, Paul and Kazman, Rick},
   title = {Software Architecture in Practice (4th Edition)},
   publisher = {Addison-Wesley Professional},
   edition = {4th},
   year = {2021}
}

@inproceedings{bosch2004software,
  title={Software architecture: The next step},
  author={Bosch, Jan},
  booktitle={Proceedings of the 1st European Workshop on Software Architecture (EWSA)},
  pages={194--199},
  year={2004},
}

@article{tyree2005architecture,
  title={Architecture decisions: Demystifying architecture},
  author={Tyree, Jeff and Akerman, Art},
  journal={IEEE Software},
  volume={22},
  number={2},
  pages={19--27},
  year={2005},
  publisher={IEEE}
}

@article{tang2006asurvey,
  title = {A survey of architecture design rationale},
  journal = {Journal of Systems and Software},
  volume = {79},
  number = {12},
  pages = {1792--1804},
  year = {2006},
  author = {Antony Tang and Muhammad Ali Babar and Ian Gorton and Jun Han}
}

@article{wei2022emergent,
    title={Emergent Abilities of Large Language Models}, 
    author={Jason Wei and Yi Tay and Rishi Bommasani and Colin Raffel and Barret Zoph and Sebastian Borgeaud and Dani Yogatama and Maarten Bosma and Denny Zhou and Donald Metzler and Ed H. Chi and Tatsunori Hashimoto and Oriol Vinyals and Percy Liang and Jeff Dean and William Fedus},
    year={2022},
    journal={arXiv preprint arXiv:2206.07682},
}

@book{moran2020design,
  title={Design Rationale: Concepts, Techniques, and Use},
  author={Moran, Thomas P and Carroll, John M},
  year={2020},
  publisher={CRC Press}
}

@inproceedings{rogers2015using,
    author = {Rogers, Benjamin and Qiao, Yechen and Gung, James and Mathur, Tanmay and Burge, Janet E.},
    title = {Using Text Mining Techniques to Extract Rationale from Existing Documentation},
    booktitle = {Proceedings of the 14th Design Computing and Cognition (DCC)},
    pages={457--474},
    year = {2015}
}

@article{sahoo2024systematic,
  title={A systematic survey of prompt engineering in large language models: Techniques and applications},
  author={Sahoo, Pranab and Singh, Ayush Kumar and Saha, Sriparna and Jain, Vinija and Mondal, Samrat and Chadha, Aman},
  journal={arXiv preprint arXiv:2402.07927},
  year={2024}
}

@inproceedings{brown2020language,
  author={Brown, Tom B. and Mann, Benjamin and Ryder, Nick and Subbiah, Melanie and Kaplan, Jared and Dhariwal, Prafulla and Neelakantan, Arvind and Shyam, Pranav and Sastry, Girish and Askell, Amanda and Agarwal, Sandhini and Herbert-Voss, Ariel and Krueger, Gretchen and Henighan, Tom and Child, Rewon and Ramesh, Aditya and Ziegler, Daniel M. and Wu, Jeffrey and Winter, Clemens and Hesse, Christopher and Chen, Mark and Sigler, Eric and Litwin, Mateusz and Gray, Scott and Chess, Benjamin and Clark, Jack and Berner, Christopher and McCandlish, Sam and Radford, Alec and Sutskever, Ilya and Amodei, Dario},
  title={Language models are few-shot learners},
  booktitle = {Proceedings of the 34th Annual Conference on Neural Information Processing Systems (NeurIPS)},
  pages={457--474},
  year = {2020}
}

@article{deljouyi2024leveraging,
  title={Leveraging large language models for enhancing the understandability of generated unit tests},
  author={Deljouyi, Amirhossein and Koohestani, Roham and Izadi, Maliheh and Zaidman, Andy},
  journal={arXiv preprint arXiv:2408.11710},
  year={2024}
}

@article{wooldridge1995the, 
title={Intelligent agents: theory and practice}, 
journal={The Knowledge Engineering Review}, 
author={Wooldridge, Michael and Jennings, Nicholas R.}, 
year={1995}, 
}

@article{bairi2024codeplan,
  title={Codeplan: Repository-level coding using llms and planning},
  author={Bairi, Ramakrishna and Sonwane, Atharv and Kanade, Aditya and Iyer, Arun and Parthasarathy, Suresh and Rajamani, Sriram and Ashok, B and Shet, Shashank},
  journal={Proceedings of the ACM on Software Engineering},
  volume={1},
  number={FSE},
  pages={675--698},
  year={2024},
  publisher={ACM}
}

@inproceedings{hu2023large,
  author={Hu, Sihao and Huang, Tiansheng and İlhan, Fatih and Tekin, Selim Furkan and Liu, Ling},
  booktitle={Proceedings of 5th IEEE International Conference on Trust, Privacy and Security in Intelligent Systems and Applications (TPS-ISA)}, 
  title={Large Language Model-Powered Smart Contract Vulnerability Detection: New Perspectives}, 
  year={2023},
  pages={297--306},
}

@inproceedings{jansen2005software,
  author={Anton Jansen and Jan Bosch},
  booktitle={Proceedings of 5th Working IEEE/IFIP Conference on Software Architecture (WICSA)}, 
  title={Software Architecture as a Set of Architectural Design Decisions}, 
  year={2005},
  pages={109--120}
}

@article{davide2013value,
  author = {Falessi, Davide and Briand, Lionel C. and Cantone, Giovanni and Capilla, Rafael and Kruchten, Philippe},
  title = {The value of design rationale information},
  year = {2013},
  journal = {ACM Transactions on Software Engineering and Methodology},
  volume={22},
  number={3},
  pages={1--32}
}

@inproceedings{balloccu2024leak,
  author={Balloccu, Simone  and Schmidtov{\'a}, Patr{\'i}cia  and Lango, Mateusz  and Dusek, Ondrej},
  booktitle={Proceedings of the 18th Conference of the European Chapter of the Association for Computational Linguistics (EACL)}, 
  title={Leak, Cheat, Repeat: Data Contamination and Evaluation Malpractices in Closed-Source LLMs}, 
  year={2024},
  pages={67--93}
}

@article{tang2007rationale,
title = {A rationale-based architecture model for design traceability and reasoning},
journal = {Journal of Systems and Software},
volume = {80},
number = {6},
pages = {918--934},
year = {2007},
author = {Antony Tang and Yan Jin and Jun Han}
}

@inproceedings{davide2008value,
  author={Falessi, Davide and Cantone, Giovanni and Kruchten, Philippe},
  booktitle={Proceedings of 7th Working IEEE/IFIP Conference on Software Architecture (WICSA)}, 
  title={Value-Based Design Decision Rationale Documentation: Principles and Empirical Feasibility Study}, 
  year={2008},
  pages={189--198},
}

@inbook{vanderVen2006design,
author={Jan Salvador van der Ven and Anton Jansen and Jos A. G. Nijhuis and Jan Bosch},
title={Design Decisions: The Bridge between Rationale and Architecture},
bookTitle={Rationale Management in Software Engineering},
year={2006},
publisher={Springer Berlin Heidelberg},
pages={329--348}
}

@inproceedings{soliman2024exploring,
author = {Soliman, Mohamed},
title = {Exploring Architectural Design Decisions in Mailing Lists and Their Traceability to Issue Trackers},
year = {2024},
booktitle = {Proceedings of 18th European Conference on Software Architecture (ECSA)},
pages = {307--323},
}

@article{xie2024leveraging,
  title={How Effective are Large Language Models in Generating Software Specifications?},
  author={Danning Xie and Byungwoo Yoo and Nan Jiang and Mijung Kim and Lin Tan and Xiangyu Zhang and Judy S. Lee},
  journal={arXiv preprint arXiv:2306.03324},
  year={2024}
}

@inproceedings{nam2024using,
author = {Nam, Daye and Macvean, Andrew and Hellendoorn, Vincent and Vasilescu, Bogdan and Myers, Brad},
title = {Using an LLM to Help With Code Understanding},
year = {2024},
booktitle = {Proceedings of 46th IEEE/ACM International Conference on Software Engineering (ICSE)},
pages = {1184--1196}
}

@article{shaik2024s3s3llm,
  title={S3LLM: Large-Scale Scientific Software Understanding with LLMs using Source, Metadata, and Document},
  author={Kareem Shaik and Dali Wang and Weijian Zheng and Qinglei Cao and Heng Fan and Peter Schwartz and Yunhe Feng},
  journal={arXiv preprint arXiv:2403.10588},
  year={2024}
}

@article{pan2025codellmsunderstanddesign,
  title={Do Code LLMs Understand Design Patterns?},
  author={Zhenyu Pan and Xuefeng Song and Yunkun Wang and Rongyu Cao and Binhua Li and Yongbin Li and Han Liu},
  journal={arXiv preprint arXiv:2501.04835},
  year={2025}
}

@inproceedings{soliman2025do,
author = {Soliman, Mohamed and Keim, Jan},
year = {2025},
title = {Do Large Language Models Contain Software Architectural Knowledge? An Exploratory Case Study with GPT},
booktitle   = {Proceedings of 22nd IEEE International Conference on Software Architecture (ICSA)},
pages={13--24}
}

@inproceedings{capilla2008effort,
  author={Capilla, Rafael and Nava, Francisco and Carrillo, Carlos},
  booktitle={Proceedings of 23rd IEEE/ACM International Conference on Automated Software Engineering (ASE)}, 
  title={Effort Estimation in Capturing Architectural Knowledge}, 
  year={2008},
  pages={208--217},
}

@article{landis1977theMO,
  title={The measurement of observer agreement for categorical data.},
  author={J Richard Landis and Gary G. Koch},
  journal={Biometrics},
  year={1977},
  pages={159--74},
  volume = {33},
  number = {1}
}

@article{liu2024large,
  title={Large Language Model-Based Agents for Software Engineering: A Survey},
  author={Junwei Liu and Kaixin Wang and Yixuan Chen and Xin Peng and Zhenpeng Chen and Lingming Zhang and Yiling Lou},
  journal={arXiv preprint arXiv:2409.02977},
  year={2024} 
}

@inproceedings{mahajan2020recommending,
author = {Mahajan, Sonal and Abolhassani, Negarsadat and Prasad, Mukul R.},
title = {Recommending stack overflow posts for fixing runtime exceptions using failure scenario matching},
year = {2020},
booktitle = {Proceedings of the 28th ACM Joint Meeting on European Software Engineering Conference and Symposium on the Foundations of Software Engineering (FSE)},
pages = {1052--1064},
}

@inproceedings{lin2004rouge,
  title={Rouge: A package for automatic evaluation of summaries},
  author={Lin, Chin-Yew},
  booktitle={Proceedings of the 42nd annual meeting of the Association for Computational Linguistics (ACL)},
  pages={74--81},
  year={2004}
}

@inproceedings{papineni2002bleu,
  author = {Papineni, Kishore and Roukos, Salim and Ward, Todd and Zhu, Wei-Jing},
  title = {BLEU: a method for automatic evaluation of machine translation},
  year = {2002},
  booktitle={Proceedings of the 40th annual meeting of the Association for Computational Linguistics (ACL)},
  pages={311--318}
}

@inproceedings{banerjee2005meteor,
  title={METEOR: An automatic metric for MT evaluation with improved correlation with human judgments},
  author={Banerjee, Satanjeev and Lavie, Alon},
  booktitle={Proceedings of the 43rd annual meeting of the Association for Computational Linguistics (ACL)},
  pages={65--72},
  year={2005}
}

@inproceedings{guerra2025acccessing,
  title={Accessing LLMs for Front-end Software Architecture Knowledge},
  author={Luiz Pedro Franciscatto Guerra and Neil Ernst},
  booktitle={Proceedings of the 2nd International Workshop on Designing Software (Designing)},
  year={2025}
}

@inproceedings{casillo2025towards,
    author = {Casillo, Francesco and Mastropaolo, Antonio and Bavota, Gabriele and Deufemia, Vincenzo and Gravino, Carmine},
    title = {Towards Generating the Rationale for Code Changes},
    booktitle = {Proceedings of 33rd International Conference on Program Comprehension (ICPC), RENE Track},
    pages = {327--338},
    year = {2025}
}

@mastersthesis{chaudron2024exploring,
  author  = {Floris M.J. Pex},
  title   = {Exploring Software Explanation using Retrieval Augmented Generation},
  school  = {Eindhoven University of Technology},
  year    = {2024}
}

@article{jansen2008documenting,
title = {Documenting after the fact: Recovering architectural design decisions},
journal = {Journal of Systems and Software},
volume = {81},
number = {4},
pages = {536--557},
year = {2008},
author = {Anton Jansen and Jan Bosch and Paris Avgeriou},
}

@inproceedings{arman2018recovering,
  author={Shahbazian, Arman and Kyu Lee, Youn and Le, Duc and Brun, Yuriy and Medvidovic, Nenad},
  booktitle={Proceedings of 15th IEEE International Conference on Software Architecture (ICSA)}, 
  title={Recovering Architectural Design Decisions}, 
  pages = {95--104},
  year={2018},
}

@article{runeson2009guidelines,
  title={Guidelines for conducting and reporting case study research in software engineering},
  author={Runeson, Per and H{\"o}st, Martin},
  journal={Empirical Software Engineering},
  volume={14},
  pages={131--164},
  year={2009},
  publisher={Springer},
  number={2}
}

@article{israel1992determining,
  title={Determining sample size},
  author={Israel, Glenn D and others},
  year={1992},
  publisher={University of Florida Cooperative Extension Service, Institute of Food and Agriculture Sciences}
}

@book{Wohlin2012ESE,
   author = {Wohlin, Claes and Runeson, Per and Höst, Martin and Ohlsson, Magnus C and Regnell, Björn and Wesslén, Anders},
   title = {{Experimentation in Software Engineering}},
   publisher = {Springer Science \& Business Media},
   year = {2012}
}

@inproceedings{stol2016grounded,
author = {Stol, Klaas-Jan and Ralph, Paul and Fitzgerald, Brian},
title = {Grounded theory in software engineering research: a critical review and guidelines},
year = {2016},
pages = {120--131},
booktitle={Proceedings of the 38th International conference on software engineering (ICSE)},
publisher={ACM}
}

@inproceedings{shashin2014do,
author = {Shahin, Mojtaba and Liang, Peng and Li, Zengyang},
title = {Do architectural design decisions improve the understanding of software architecture? two controlled experiments},
year = {2014},
publisher = {ACM},
booktitle = {Proceedings of the 22nd International Conference on Program Comprehension (ICPC)},
pages = {3--13},
}

@article{dhar2025drafting,
  title={DRAFT-ing Architectural Design Decisions using LLMs},
  author={Rudra, Dhar and Adyansh, Kakran and Amey, Karan and Karthik, Vaidhyanathan and Vasudeva, Varma},
  journal={arXiv preprint arXiv:2504.08207},
  year={2025} 
}

@inproceedings{díaz-pace2024helping,
author={D{\'i}az-Pace, J. Andr{\'e}s and Tommasel, Antonela
and Capilla, Rafael},
title={Helping Novice Architects to Make Quality Design Decisions Using an LLM-Based Assistant},
booktitle={Proceedings of the 18th European Conference on Software Architecture (ECSA)},
year={2024},
publisher={Springer},
pages={356--374}
}

@inproceedings{shahin2009architectural,
  title={Architectural Design Decision: Existing Models and Tools},
  author={Shahin, Mojtaba and Liang, Peng and Khayyambashi, Mohammad Reza},
  booktitle={Proceedings of the Joint Working IEEE/IFIP Conference on Software Architecture \& European Conference on Software Architecture (WICSA/ECSA)},
  pages={293--296},
  year={2009},
  organization={IEEE}
}

%%
% %% If your work has an appendix, this is the place to put it.
% \appendix









\end{document}